\documentclass[12pt]{article}

\usepackage{epsfig}
\usepackage{amssymb}
\usepackage{graphicx}
\usepackage{color}
\usepackage{subfigure}
\usepackage[T1]{fontenc}
\usepackage{jheppub} 

\begin{document}

\baselineskip 6mm
\renewcommand{\thefootnote}{\fnsymbol{footnote}}


\newcommand{\nc}{\newcommand}
\newcommand{\rnc}{\renewcommand}



\newcommand{\tcb}{\textcolor{blue}}
\newcommand{\tcr}{\textcolor{red}}
\newcommand{\tcg}{\textcolor{green}}


\def\be{\begin{eqnarray}}
\def\ee{\end{eqnarray}}
\def\nn{\nonumber\\}


\def\ct{\cite}
\def\la{\label}
\def\eq#1{(\ref{#1})}


\def\a{\alpha}
\def\b{\beta}
\def\g{\gamma}
\def\G{\Gamma}
\def\d{\delta}
\def\D{\Delta}
\def\e{\epsilon}
\def\et{\eta}
\def\ph{\phi}
\def\Ph{\Phi}
\def\ps{\psi}
\def\Ps{\Psi}
\def\k{\kappa}
\def\l{\lambda}
\def\L{\Lambda}
\def\m{\mu}
\def\n{\nu}
\def\th{\theta}
\def\Th{\Theta}
\def\r{\rho}
\def\s{\sigma}
\def\S{\Sigma}
\def\ta{\tau}
\def\o{\omega}
\def\O{\Omega}
\def\pr{\prime}


\def\half{\frac{1}{2}}

\def\goto{\rightarrow}

\def\na{\nabla}
\def\grad{\nabla}
\def\curl{\nabla\times}
\def\div{\nabla\cdot}
\def\pa{\partial}
\def\fr{\frac}

\def\bra{\left\langle}
\def\ket{\right\rangle}
\def\lb{\left[}
\def\lc{\left\{}
\def\ls{\left(}
\def\lp{\left.}
\def\rp{\right.}
\def\rb{\right]}
\def\rc{\right\}}
\def\rs{\right)}

\def\vac#1{\mid #1 \rangle}


\def\td#1{\tilde{#1}}
\def\check{ \maltese {\bf Check!}}


\def\Tr{{\rm Tr}\,}
\def\det{{\rm det}}
\def\text#1{{\rm #1}}


\def\bc#1{\nnindent {\bf $\bullet$ #1} \\ }
\def\ch {$<Check!>$ }
\def\ss {\vspace{1.5cm}}
\def\inf{\infty}

\begin{titlepage}

\hfill\parbox{5cm} { }

\vspace{25mm}

\begin{center}
{\Large \bf Holographic two-point functions in medium}

\vskip 1. cm
Chanyong Park$^a$\footnote{e-mail : cyong21@gist.ac.kr},
Se-Jin Kim $^b$\footnote{e-mail : sejin817@kookmin.ac.kr}
and Jung Hun Lee$^b$\footnote{e-mail : junghun.lee@kookmin.ac.kr}

\vskip 0.5cm

{\it  $^a\,$Department of Physics and Photon Science, Gwangju Institute of Science and Technology, Gwangju 61005, Korea}\\

{\it $^b\,$College of General Education, Kookmin University, Seoul, 02707, Korea} \\

\end{center}

\thispagestyle{empty}

\vskip1cm


\centerline{\bf ABSTRACT} \vskip 4mm

We study two-point correlation function in a medium composed of two kinds of matter, which is the dual of a three-dimensional generalized $p$-brane gas geometry. Following the holographic prescription, we calculate temporal and spatial two-point functions in the medium. In general, the screening effect of the medium makes two-point functions decrease more rapidly than the CFT's two-point function. In the extremal limit, however, we find that a temporal two-point function is still conformal. This indicates that a two-dimensional UV CFT flows into a one-dimensional quantum mechanics in the IR limit. This is consistent with the fact that the near horizon geometry in the extremal limit reduces to AdS$_2$. We also investigate holographic mutual information representing the correlation between two subsystems. We show that a critical distance in the IR region, where the mutual information vanishes, leads to a similar behavior to the correlation length of a two-point function.

\vspace{2cm}


\end{titlepage}

\renewcommand{\thefootnote}{\arabic{footnote}}
\setcounter{footnote}{0}



\section{Introduction}

For a long time, theoretical physics has tried to find a theory that reconciles general relativity and quantum mechanics. After the proposal of the AdS/CFT correspondence (or holography)  which relates a $(d+1)$-dimensional gravity theory in an Anti-de Sitter (AdS)  to a $d$-dimensional conformal field theory (CFT) \cite{Maldacena:1997re,Witten:1998qj,Gubser:1998bc,Maldacena:2002mn}, the holographic method has been widely used in investigating strongly interacting quantum field theory (QFT) via its dual gravity theory \cite{Hartnoll:2008kx,Maldacena:2012xp,Myers:2012ed,Bueno:2015rda,Kim:2018mgz,Koh:2018rsw,Park:2018ebm,Cooper:2018cmb,Almheiri:2019hni}. Due to the large symmetry of an AdS space and CFT, their holographic relation is well understood. However, if we consider a nonconformal field theory, it is usually hard to find its dual gravity theory because of the absence of sufficient symmetry. If the dual gravity is known,
the holography is applied to a variety of strongly interacting systems like the condensed matter physics, quantum information theory and nuclear physics \cite{Policastro:2001yc,Hartnoll:2008vx,Swingle:2009bg,Karch:2010mn,Nozaki:2012zj,Erdmenger:2015spo,Lashkari:2015hha,Park:2017ray,Akutagawa:2020yeo,Shenker:2013pqa}. In this work, we take into account a generalized $p$-brane gas geometry whose dual field theory corresponds to a medium composed of several matters. Since a medium generally has a nontrivial ground state, it is important to understand the physical properties of a medium. However, it is not easy to study a medium because it requires a nonperturbative method. Here, we investigate various physical properties of a medium by applying the holographic technique.

Following the AdS/CFT correspondence, various non-local observables of QFTs such as a Wilson loop and entanglement entropy can be realized as the geometric objects extending to the dual geometry \cite{Maldacena:1998im,Ryu:2006bv,Ryu:2006ef,Hubeny:2007xt,Casini:2011kv,Nishioka:2009un,Takayanagi:2012kg,Kim:2012tu}. This relation was further generalized into a two-point correlation function of local operators. It was argued that a two-point function of a boundary QFT maps to a geodesic length connecting two boundary operators. For a medium composed of strongly interacting matters, the well-established perturbative calculating a two-point function is not valid anymore. In this situation, the AdS/CFT correspondence can shed light on finding a nonperturbative two-point function. Since a nonperturbative two-point function is affected by the background medium, it may give us information about the medium effect of a nontrivial ground state. 

In the present work, we take into account a $p$-brane gas geometry which is the dual of a medium.  At the boundary of a $p$-brane gas geometry, $p$-branes maps to a $(p-1)$-dimensional objects \cite{Chakrabortty:2011sp}. For $p=1$, $p$-branes become open strings and their ends at the boundary corresponds to massive particles. This $p$-brane gas geometry for $p=1$ was known as the string cloud geometry which provides a useful playground to understand various physical properties of quark-gluon plasma \cite{Lee:2009bya,Park:2017ray} and thermal states of heavy nuclei \cite{Chakrabortty:2011sp,Chakrabortty:2016xcb}. For $p=0$, a $p$-brane gas geometry reduces to the well-known AdS black hole and $0$-branes maps to massless field like gluons. A $p$-brane gas geometry further generalizes into a generalized $p$-brane gas geometry including several different types of $p$-branes, which also satisfies the bulk equations of motion. A generalized $p$-brane gas geometry resembles a black hole with several hairs which are associated with several different types of matters on the dual QFT. For a specific parameter region, intriguingly, a generalized $p$-brane gas geometry allows an extremal limit which corresponds to a ground state of a medium. On this background geometry, we holographically investigate some physical quantities of a medium like the entanglement entropy, drag force, two-point functions and mutual information.

In the holographic entanglement entropy study \cite{Bhattacharya:2012mi,Kim:2015rvu,Park:2020nvo,Park:2022fqy}, it was well known that an entanglement temperature shows a universal feature. For an AdS black hole, an entanglement temperature is inversely proportional to a subsystem size regardless of the dimension and shape of a subsystem \cite{Bhattacharya:2012mi,Park:2015afa}. This is a typical feature of a relativistic QFT. For a Lifshitz geometry whose dual QFT is nonrelativistic, it was shown that an entanglement temperature shows a different universality. Then, it would be interesting to ask what kind of universality appears in a medium which contains massive objects. By applying the holographic method, we find that an entanglement temperature in a medium consisting of $(p-1)$-dimensional objects is universally proportional to $l^{p-1}$ where $l$ is a subsystem size.

It is important to know two-point function of a local operator to understand dynamics and spectrum of a local operator \cite{Park:2021tpz,Park:2022mxj,Georgiou:2022ekc}. In a medium, however, it is not easy to calculate such a two-point function including all quantum corrections. In this work, we look into nonperturbative two-point functions in a two-dimensional medium. When we consider a medium composed of massless and massive particles, its dual gravity can be described by a three-dimensional generalized $p$-brane gas geometry with $p=0$ and $1$. This generalized $p$-brane gas geometry resembles a black hole having two hairs, so that it can permit an extremal limit which corresponds to the zero temperature limit of the dual QFT. For a non-extremal limit, we show that temporal and spatial two-point functions in the IR limit (with a large time interval and large distance) suppress exponentially, as expected, due to the screening effect of a background medium. However, two-point functions show different behaviors In the extremal limit. Although a spatial two-point function in the IR limit suppresses exponentially, a temporal two-point function decays by a power law similar to the CFT case. As a result, the resulting IR theory becomes a one-dimensional one because of exponential suppression of the spatial correlation. This means that a two-dimensional UV CFT in a medium can flows into a one-dimensional conformal quantum mechanics. This is consistent with that fact that the near horizon geometry of an extremal black hole is given by AdS$_2 \times R$ where the spatial direction $R$ decouples.

Lastly, we study the mutual information between two macroscopic subsystems to see the connection to the microscopic two-point function. Since the mutual information may be associated with the collective behavior of microscopic spatial correlation functions, we may expect that the correlation function in the large distance limit provide a main contribution to the mutual information in the large subsystem size limit. By applying the holographic method, we compare a spatial two-point function with a mutual information. After defining a critical distance where a mutual information vanishes, we show that this critical distance behaves in a similar way to the correlation length of a spatial two-point function defined in the large distance limit.


We have organized our paper as follows. In Sec. \ref{sec:2}, we briefly review a $p$-brane gas geometry whose dual field theory describes a medium consisting of $(p-1)$-dimensional objects. Using this $p$-brane gas geometry, we discuss the universal feature of the entanglement temperature relying on the value of $p$. In Sec. \ref{sec:3}, we take into account a generalized $p$-brane gas geometry composed of two kinds of $p$-branes. A generalized $p$-brane gas geometry resembles a black hole with two hairs corresponding to the densities of $p$-branes. From this black hole geometry, we investigate thermodynamic properties of the dual field theory. We further study the medium effect on drag force and two-point functions in the medium. In Sec. \ref{sec:4}, we further study the correlation between two macroscopic subsystems by calculating the mutual information. We show that a critical distance of the mutual information in the IR region behaves similar to the correlation length of a spatial two-point function. Lastly, we close this work with some concluding remarks in Sec. \ref{sec:5}.



\section{ $p$-brane gas geometry and its holographic dual \label{sec:2}}

To understand strongly interacting systems holographically, we have to  first know its dual gravity. It was known that a $(d+1)$-dimensional AdS geometry maps to the vacuum of a $d$-dimensional CFT. To further study a nontrivial ground state in medium, we also take into account appropriate bulk fields. Assuming that $p$-branes extending to the radial direction are uniformly distributed in an AdS space. It was known that they correspond to $(p-1)$-dimensional extended objects on the dual QFT side \cite{Park:2020jio,Park:2021wep}. The gravitational backreaction of a $p$-brane gas modifies the background AdS space and results in the $p$-brane gas geometry which is dual of the medium we will consider. On the gravity side, more precisely, the $p$-brane gas geometry is described by the following action \ct{Park:2021wep}
\be
S = \fr{1}{2 \k^2} \int d^{d+1} x \sqrt{- G} \ls {\cal R} - 2 \L \rs  + T_p  N_p  \int d^{p+1} \xi \sqrt{-h} \  \pa^{\a} x_{M}  \, h_{\a\b} \, \pa^{\b} x_{N} \, G^{MN}  ,
\ee
where $N_p$ and $T_p$ are the number and tension of $p$-branes, respectively. Here, $d \ge p$ and $\xi^\a$ indicate the coordinate of the $p$-brane's worldvolume.

After varying the action (see the details in Ref. \cite{Park:2020jio,Park:2021wep}), we find the following Einstein equation 
\be
R_{MN} - \half G_{MN} R + G_{MN} \L = 2 \k^2 T_{MN} .  \la{Equation:Einstein}
\ee
where the stress tensor of $p$-branes is given by
\be
T_{MM}  = - \fr{T_p n_p \, z^{d-p} }{R^{d-p}} \lc G_{tt} ,  \fr{p-1}{d-1} G_{11} , \cdots,    \fr{p-1}{d-1}  G_{(d-1) \, (d-1)}  , G_{zz} \rc  .
\ee
Solving this equation leads to the following black hole-like solution
\begin{equation}
ds^2 = \frac{R^2}{z^2} \left(-f(z) dt^2  + \frac{1}{f(z)} dz^2 + \d_{ij} dx^i dx^j\right),  \la{Ansatz:metricform}
\end{equation}
with a nontrivial blackening factor 
\be
f (z) &=& 1  -   \fr{z^{d-p}}{z_h^{d-p}}  .
\ee
The horizon $z_h$ is determined as a function of the $p$-brane's number density, $n_p$,
\be
z_h^{d-p} =  \fr{1}{c_p  \, \k^2 n_p \, T_p \, R^{p+2-d} } ,   \la{Relation:hodensity}
\ee
where $c_p$ is an appropriate numerical value, for example, $c_1 = 4/3$, $c_2 = 2/3$, and $c_3 = 4/9$ for $d=4$. This geometric solution was called the $p$-brane gas geometry \cite{Park:2020jio,Park:2021wep} and is the dual of a medium composed of $(p-1)$-dimensional solitonic objects on the dual QFT side.

Due to the existence of a horizon in the $p$-brane gas geometry, the Hawking temperature and Bekenstein-Hawking entropy are well defined as
\be
T_H &=& \fr{d-p}{4 \pi z_h} , \nn
\fr{S_{BH}}{V} &=& \fr{2 \pi }{\k^2} \fr{R^{d-1} }{z_h^{d-1} } ,
\ee
where $V$ is an appropriately regularized volume. Using the thermodynamic law, the internal energy of the system is given by
\be \la{internalE}
\fr{E}{V} = \frac{(d-1) (d-p)    }{2 d \, \kappa ^2 } \fr{R^{d-1}}{z_h^{d}}  \sim n_p^{d/(d-p)} .
\ee
This result indicates that the energy density is proportional to the density of $p$-branes with the power given by $d/(d-p)$. 

Now, we investigate thermodynamics of a $p$-brane geometry which is the dual of a nontrivial ground state in a medium. In the holographic study, one can define entanglement temperature as representing the ratio of the entanglement entropy to the subsystem's energy. Unlike the temperature of thermal systems, the entanglement temperature is not globally defined because it crucially depends on the subsystem size. Nevertheless, it was known that the entanglement temperature shows a universal feature in a UV region. For the dual QFT of an ordinary AdS black hole, for instance, the entanglement temperature is inversely proportional to the subsystem size \cite{Bhattacharya:2012mi,Kim:2015rvu}. This is a typical feature of relativistic QFTs. If we further consider a medium consisting of nonrelativistic massive objects, it would be interesting to ask how the ground state entanglement entropy increases when the ground state energy varies? The answer to this question is related to the universal feature of the entanglement entropy in a medium. To discuss this feature, from now on,
we focus on the entanglement entropy in a strip-shaped region for $d=4$. This work easily generalizes to other dimensions. We have checked that the universality we will study below generally appears regardless of the shape of the entangling region. 

Let us first parameterize a subsystem in a strip-shaped region as
\be
- \fr{l}{2} \le x\equiv x_1 \le \fr{l}{2}  \quad {\rm and} \quad  - \fr{L}{2} \le x_2, x_3 \le \fr{L}{2} ,
\ee
where $L$ is an appropriate regulator and $l < L$ corresponds to the width of the strip. Then, the holographic entanglement entropy is determined by
\be
S_E = \fr{R^3 L^2 }{4 G}  \int_{-l/2}^{l/2} d x \, \fr{ \sqrt{f + z'^2}}{z^3 \, \sqrt{f}} .
\ee
Due to the invariance under $x \to -x$, there exists a turning point $z_0$ and a minimal surface extends only to $0 \le z \le z_0$. Above a translational symmetry in the $x$-direction leads to a conserved quantity. In the UV region with $z_0 \ll z_h$, the conserved quantity leads to the following relation
\be
l &=&2  \int_0^{z_0} du \ \frac{ z^3}{\sqrt{f} \sqrt{z_0^6-z^6}}   \la{Formula:subsystemsize}
=2\int^{z_0}_0 dz  \frac{z^3}{\sqrt{z_0^6-z^6}}\sum_{n=0}^\infty\frac{(2n)!}{4^n(n!)^2}\biggl(\frac{z}{z_h}\biggr)^{(4-p)n} .
\ee
 In this case the turning point $z_0$ determines the subsystem size $l$ which can be reinterpreted as the inverse of the energy scale. Therefore, the case of $z_0/z_h \ll 1$ leads to a small subsystem size which corresponds to a high or UV energy region. After performing the integration, we can determine  $z_0$ as a function of $l$
\be
z_0=\frac{2\,\Gamma({\scriptstyle\frac{7}{6}})}{\sqrt{\pi}\Gamma({\scriptstyle\frac{5}{3}})}l
-\frac{2^{5-p}}
{3\pi^{(5-p)/2}}\frac{\Gamma({\scriptstyle\frac{4}{3}-\frac{p}{6}})}{\Gamma({\scriptstyle\frac{11}{6}-\frac{p}{6}})}\biggl(\frac{\Gamma({\scriptstyle\frac{7}{6}})}{\Gamma({\scriptstyle\frac{5}{3}})}\biggr)^{6-p}\biggl(\frac{l}{z_h}\biggr)^{4-p}l+\cdots .
\ee

Similarly, the entanglement entropy in the UV region allows the following perturbative expansion
\be
S_E
&=&\frac{R^3L^2}{2G}\int^{z_0}_{\epsilon_{uv}} dz\frac{z_0^3}{z^3\sqrt{z_0^6-z^6}}\sum_{n=0}^\infty\frac{(2n)!}{4^n(n!)^2}\biggl(\frac{z}{z_h}\biggr)^{(4-p)n},
\ee
where we introduce a UV cutoff $\epsilon_{uv}$ to regularize UV divergences.  The leading contribution to the UV entanglement entropy comes from $n=0$ 
\be
S_E&=&\frac{R^3}{4G}\biggl[\frac{L^2}{\epsilon_{uv}^2}
-4\pi^{3/2}\biggl(\frac{\Gamma({\scriptstyle\frac{2}{3}})}{\Gamma({\scriptstyle\frac{1}{6}})}\biggr)^3\frac{L^2}{l^2}\biggr] .
\ee
This leading contribution corresponds to that of a four-dimensional CFT. On the other hand, the medium effect appears at the first correction for $n=1$. Depending on the background matter, the UV entanglement entropy shows the following different first corrections;
\begin{itemize}
\item  for $p=0, 1$
\be \la{resultp01}
\Delta S_E=\frac{R^3}{4G}\biggl[
\frac{\pi^{(p-1)/2}}{3\times2^{4-p}}\frac{\Gamma({\scriptstyle\frac{1}{3}-\frac{p}{6}})}{\Gamma({\scriptstyle\frac{11}{6}-\frac{p}{6}})}\biggl(\frac{\Gamma({\scriptstyle\frac{2}{3}})}{\Gamma({\scriptstyle\frac{1}{6}})}\biggr)^{p-2} \ 
\frac{L^2 \  l^{2-p}}{z_h^{4-p}}  +\cdots
\biggr].
\ee
\item  for $p=2$
\be \la{resultp2}
\Delta S_E=\frac{R^3}{4G}\biggl[
\biggl(\,
\log\frac{l}{\epsilon_{uv}}
+\log\biggl(\frac{\Gamma({\scriptstyle\frac{1}{6}})}{2^{2/3}\sqrt{\pi}\Gamma({\scriptstyle\frac{2}{3}})}\biggr)
-\frac{1}{3}\,\biggr) \
\frac{L^2}{z_h^2} +\cdots\biggl].
\ee
\item  for $p=3$
\be \la{resultp3}
\Delta S_E=\frac{R^3}{4G}\biggl[
\biggl(\,
\frac{l}{\epsilon_{uv}}
-\frac{2^{8/3}\pi^{5/2}}{\Gamma({\scriptstyle\frac{1}{6}})^3}
\,\biggr) \  
\frac{L^2}{z_h \ l} +\cdots
\biggr].
\ee
\end{itemize}
This result shows that for $p \ge 2$ there are additional UV divergences caused by the matter. Removing such additional UV divergences with an appropriate renormalization procedure, the background matter gives rise to the finite first correction 
\be
\D S_E \sim l^{2 -p} .
\ee

Recalling that the vacuum energy of a CFT is zero, the internal energy in \eq{internalE} corresponds to that of the $(p-1)$-dimensional objects on the dual field theory side
\be \label{delE}
\Delta E=\frac{3(4-p)}{8 \k^2 R}\frac{R^4}{z_h^4} V,
\ee
where $V = L^2 \, l$ is the regularized volume of a subsystem. Comparing the internal energy and entanglement entropy, we find that the entanglement temperature in the UV region is given by
\be
T_{E} = \fr{\D E}{\D S_E} \sim  \  n_p^{p/(4-p)}  \  l^{p-1}  .  \la{Result:univET}
\ee
For $p=0$, the $p$-brane gas geometry reduces to an ordinary AdS black hole. In this case, the dual matter is massless adjoint fields which are relativistic. Therefore, the entanglement temperature evaluated in the $0$-brane gas geometry follows  the known universality, $T_E \sim l^{-1}$ \cite{Bhattacharya:2012mi}. However, the $p$-brane gas with $p \ne 0$ shows a different universality, $T_E \sim l^{p-1}$. This is because $p$-branes for $p \ne 0$ corresponds to nonrelativistic massive objects. As a result, the above result shows that the entanglement temperature of a nonrelativistic QFT reveals a different universality depending on the matter.

In the IR regime ($l \to \infty$), the entanglement entropy can be rewritten as
\be
S_E = \fr{ R^3 L^2\,  l }{4 G z_0^3 } + \fr{R^3 L^2}{2 G u_0^3} \int_0^{z_0} dz \ \fr{\sqrt{z_0^6 - z^6}}{z^3 \sqrt{f}} .
\ee
Here, the second term contains a UV divergence, $1/\e^2$ at $\e \to 0$. After renormalizing such a UV divergence, the main contribution to the IR entanglement entropy comes from the first term. This is because the first term diverges for $l \to \infty$ but the second term is finite after the renormalization procedure. Recalling that the volume of the subsystem is given by $V = L^2 l $, this result shows that the IR entanglement entropy follows a volume law, which is equivalent to the black hole entropy.  On the dual QFT side, this implies that the leading IR entanglement entropy in the medium is given by
\be
S_E 
&=&  \fr{ V  \ls c_p  \, \k^2  R^2 T_p   \rs^{\frac{3}{4-p}}   }{4 G}  \  n_p^{\frac{3}{4-p}}   .
\ee
This result shows that the IR entanglement entropy increases with the medium density, for example, $n_0^{3/4}$ for massless field, $n_1$ for massive particles, $n_2^{3/2}$ for cosmic strings, and $n_3^3$ for domain walls.

\section{Holographic dual of a medium composed of two kinds of matter \label{sec:3}}

In the previous section, we discussed the $p$-brane gas geometry and its dual QFT consisting of one kind of matter. This study can be further generalized into a medium composed of several different matters. For convenience, we concentrate on the $d=2$ case and set $R=1$ from now on. Then, $p$ can have $0$, $1$, and $2$ due to the constraint, $p \le d$. For $p =2$, a $2$-brane gas modifies just the background cosmological constant and its dual theory is still conformal. To consider nontrivial effect of the medium, hereafter, we take into account a $p$-brane gas geometry involving only $0$- and $1$-branes. Their gravitational backreactiion is described by the following generalized $p$-brane gas geometry
\be
ds^2 = \fr{1}{z^2} \ls - f(z) dt^2 + dx^2+ \fr{1}{f(z)} dz^2   \rs ,   \la{Metric:targetspace}
\ee
with the following blackening factor
\begin{equation}		\la{res:blackholefactor2} 
	f(z) = 1  -  \r z   - M z^2  ,
\end{equation}
which is again the solution of \eq{Equation:Einstein} with the energy-momentum tensors of $0$- and $1$-branes. Two hairs, $M$ and $\r$, are proportional to the density of $0$- and $1$-branes, respectively. They generally permit two horizons, inner and outer horizons. In this case, the outer horizon is identified with an event horizon. Denoting the event horizon by $z_h$ and representing $M$ in terms of the event horizon
\be
M = \fr{1 - \r z_h}{ z_h^2},
\ee 
the blackening factor is rewritten as
\be
f (z) =  \ls 1 - \fr{z}{z_+} \rs  \ \ls 1  - \fr{z}{z_-} \rs ,  \la{Metric:itoRoots}
\ee
with two roots
\be
z_{\pm}  =  \fr{2 z_h}{ \r z_h \pm \left| 2 - \r z_h \right| } . 
\ee
For $\r z_h \le 2$, $z_+$ and  $z_-$ correspond to an event horizon and inner horizon, respectively. For $\r z_h > 2$, two roots $z_\pm$ exchange their role. Thus, $z_- $ becomes an event horizon. From now on, we focus on the case of $\r z_h \le 2$ because the same result also appears for $\r z_h > 2$.

For $\r z_h \le 2$, $z_+=z_h$ becomes an event horizon and an inner horizon $z_{in}= z_-$ is located at 
\be
z_{in} = \fr{z_h} {\r z_h- 1} .
\ee
The regularity of the geometry at the event horizon determines the Hawking temperature as the following form
\be
T_H 
 = \fr{2 -\r z_h}{4 \pi  z_h} . 
\ee
This relation shows that the horizon can be determined as a function of temperature and density
\be
z_h = \fr{2}{4 \pi T_H + \r}.
\ee 
The Bekenstein-Hawking entropy, which corresponds to a thermal entropy of the dual QFT, is given by the area at the horizon
\be
S_{BH} = \fr{V}{4 G}  \fr{1}{ z_h} =  \fr{V}{8 G} \,  \ls 4 \pi T_H + \r \rs ,   \la{Result:ThEntropy}
\ee
where $V$ is the spatial volume of a dual QFT. This shows that the thermal entropy is linearly proportional to two independent variables, $T_H$ and $\r$.
Depending on values of the hairs, the obtained black hole solution shows several different phases:

\begin{itemize}
	
	\item For $\r z_h = 2$, two horizons merge and lead to an extremal limit with $T_H = 0$.
	
	\item For $ 1 < \r z_h < 2 $, the black hole geometry has two positive horizons satisfying $z_h  < z_{in}$.
	
	\item For $\r z_h = 1$, the inner horizon is located at $z_{in}=\infty$. In this case, the effect of $M$ disappears and the resulting geometry reduces to the string cloud geometry containing only open strings.
	
	\item For $0 \le \r z_h < 1$, the inner horizon disappears because $z_{in}$ is negative. Therefore, the resulting black hole geometry has only one positive root. For $\r=0$, in particular, $z_h = - z_{in}$ and the generalized $p$-brane gas geometry reduces to the  BTZ black hole.
	
\end{itemize}

From the above temperature and entropy, we further determine other thermodynamic quantities of the dual medium. Since the generalized $p$-brane gas geometry has two conserved hairs, its thermodynamics must be described by the following thermodynamic relation
\be
d U (S_{BH}, N, V)  = T_H \, d S_{BH} + \m \, d N - P \, d V ,
\ee
where $N = \r V$ and $U$, $\m$, and $P$ indicate an internal energy, chemical potential, and pressure, respectively. The following thermodynamic equation
\be
T_H = \lp \fr{\pa U}{\pa S_{BH}} \right|_{N,V} ,
\ee
determines the internal energy to be
\be
U (S_{BH}, N, V) = \fr{G S_{BH}^2}{\pi V} - \fr{S_{BH} N}{4 \pi V} .
\ee
This internal energy gives rise to the following chemical potential and pressure
\be
\m  &=&  \lp \fr{\pa U}{\pa N} \right|_{S_{BH},V} 
= - \fr{1}{32 \pi  G} \,  \ls 4 \pi T_H + \r \rs , \nn
P &=& -  \lp \fr{\pa U}{\pa V} \right|_{S_{BH},N} 
= \fr{1}{64 \pi G} \,  \ls 4 \pi T_H + \r \rs^2.
\ee

Defining a ground state energy  $U_0$ as the internal energy at zero temperature, the internal energy can be rewritten as 
\be
U   (T_H, \r, V) = U_0 +  U_{ex} = \fr{ (16 \pi^2 T_H^2 - \r^2) V}{64 \pi G} ,
\ee
with
\be
U_0 =  - \fr{\r^2 V}{64 \pi G }    \quad {\rm and} \quad  U_{ex} =  \fr{\pi T_H^2 V}{4 G}   . \la{Result:exenergy}
\ee
The excitation energy $U_{ex}$  (or thermal energy in the IR region) satisfies the Stefan-Boltzmann law. The ground state pressure $P_0$ at zero temperature is given by 
\be
P_0  = - \fr{U_0}{V} =  \fr{\r^2 }{64 \pi G }  .
\ee
This result indicates that $U_0$ is really the ground state energy because the equation of state parameter of the ground state is given by $w_0 = P_0 V / U_0 =-1$. If we assume that the excitation on this ground state is an ideal gas, the equation of state parameter of excitation is given by
\be
w = \fr{(P-P_0) V}{(U- U_0)} =  1 + \fr{ \r}{2 \pi T_H } .
\ee
For $\r=0$, the excitation has $w=1$ which is the equation of state parameter of massless fields for a two-dimensional QFT, as expected. This further shows that $w$ increases linearly with the density of massive particles $\r$. The heat capacity of this system at a given volume becomes
\be
C_V = \lp \fr{d U}{d T_H} \right|_{V} = \fr{\pi T_H  V}{2 G}  \ge 0 .
\ee
Therefore, the dual medium of the generalized $p$-brane gas geometry is always thermodynamically stable.

\subsection{Momentum dissipation in the medium \label{sec:3.1}}

In the previous section, we studied the physical properties of the background medium containing two kinds of matter. In this section, we further investigate the medium effect by looking into particle's motion in the medium. When a particle moves in a medium, interaction with the background matter usually disturbes the particle's motion. This medium effect can be described by a drag force \cite{Gubser:2006bz,Park:2012lzs}. In the holographic setup, the drag force is reinterpreted as the transition of the particle's momentum into the string's worldsheet momentum. In order to study such a drag force holographically, we take into account a probe open string moving in the $x$-direction. We take a static gauge, $\sigma^1=t$ and $\sigma^2 =r \equiv 1/z$, where $\s^i$ is the string's worldsheet coordinate. Assuming that variation of the string's spatial position is given by
\be
d x =v \, dt+ x' dr,
\ee 
where the prime means a derivative with respect to $r$, the open string's motion is governed by the following Nambu-Goto action  
\be     \label{res:nambu}
S = \fr{1}{2 \pi \a'} \int d^2  \s  \sqrt{ - h} = \fr{1}{2 \pi \a'} \int d t  d r  \ \sqrt{ 1 - \fr{v^2}{f} + \fr{r^4 \, f}{R^4} \ x'^2} ,
\ee
where the induced metric is defined as
\be
h_{\a\b} = \pa_\a X^\m \pa_\b X^\n G_{\m\n}   , 
\ee
and $G_{\mu\nu}$ is the metric of the target space in \eq{Metric:targetspace}. In this case, the string's velocity $v$ in the $x$-direction usually decrease because of the drag force.

Since the above action is invariant under translation in the $x$-direction, the canonical momentum $\Pi$ of $x$ is conserved. Using this conserved momentum, the string's configuration is determined by
\be
x'= \frac{2 \pi  \alpha'  \Pi  R^4}{f r^2 } 
\sqrt{\fr{ f-v^2 }{ f r^4-4 \pi ^2 \alpha'^2 \Pi ^2 R^4  }} .
\ee
To define this quantity well, the inside of the square root must be positive. Since the numerator has a root at $r = r_s$ with
\be
r_s &=&\fr{\r+\sqrt{\r^2+4M(1-v^2)}}{2(1-v^2)},
\ee
the denominator must have a root at the same position. This requirement fixes the canonical momentum to be
\be
\Pi=\fr{v}{8\pi \alpha'(1-v^2)^2 R^2}\biggl( \r +\sqrt{ \r^2 +4 M(1-v^2)}\biggr)^2.
\ee

The end of the open string corresponds to a massive particle moving with velocity $v$ at the boundary. This particle's  momentum can flow into the string worldsheet which reduces the particle's momentum. Therefore, the change of the particle's momentum is given by
\be
\D p_x = \fr{dp_x}{dt} \D t =  - \fr{1}{2 \pi \a'} \int dt \ \sqrt{-h} \  G_{x \m} h^{r\a}  \pa_\a x^{\m}   .
\ee
Near the boundary, we can see that the drag force is given by
\be
\fr{dp_x}{dt}=-\Pi.
\ee
Assuming that the particle with mass $m_0$ moves relativistically, the particle's velocity is rewritten in terms of the momentum, $v=p_x/\sqrt{m_0^2+p_x^2}$. Then, the drag force results in
\be \label{res:dragforce}
\fr{dp_x}{dt}&=&-\fr{ p_x (m_0^2+p_x^2)^{3/2}}{8\pi\alpha' R^2 \, m_0^4}
\ls \r+ \fr{\sqrt{16  \pi^2 T_H^2 m_0^2 + \r^2  p_x^2}}{\sqrt{m_0^2+p_x^2}} \rs^2  < 0.
\ee
The drag force is always negative, so that the particle's momentum monotonically decreases with time.  


First, we consider the relativistic case satisfying $p_x \gg m_0$. In the relativistic limit, the drag force is approximately
\be
\fr{dp_x}{dt} = -\frac{\rho ^2 }{2 \pi  \alpha'   R^2} \ls \fr{p_x}{m_0}\rs^4 -\frac{8 \pi ^2 T_H^2+ \rho ^2 }{2\pi  \alpha'  R^2} \ls \fr{p_x}{m_0}\rs^2 + \cdots .
\ee
For $\r \ne 0$, this result shows that the first term gives rise to the leading contribution. For $\r=0$, however, the leading contribution comes from the second term. Assuming that the initial momentum is  $p_0$ at $t=0$, the momentum for $\r \ne 0$ becomes
\be
p_x (t) = p_0-\frac{\rho ^2 }{2 \pi  \alpha'   R^2}  \ls \fr{p_0 }{m_0} \rs^4 t + {\cal O} (t^2) .
\ee
Here $p_0/m_0 \gg 1$ because $p_0 \ge p_x (t)$. Therefore, the particle's momentum rapidly decrease when the medium contains massive fundamental particles. Note that this result is valid only in the early time because the rapid reduction of the momentum for $\r \ne 0$ makes a relativistic particle nonrelativisitic. For $\r=0$, on the other hand, the momentum decreases by
\be
p_x (t) = p_0 - \frac{2 \pi  T_H^2}{\alpha'   R^2}   \ls \fr{p_0 }{m_0}\rs^2  t + {\cal O} (t^2) .
\ee
In this case, the momentum also decreases rapidly but much slower than the case of $\r \ne 0$. As a result, the drag force is much stronger in medium composed of massive particles than massless particles.


Now, we move to the non-relativistic case satisfying $p_x\ll m_0$. Even for the previous relativistic case, the drag force makes a relativistic particle's motion nonrelativisitic, after sufficient time evolution. Therefore, the particle's late time behavior is always described by a nonrelativistic motion. In this nonrelativisitic limit, the leading drag force is given by
\be
\fr{dp_x}{dt} =  -\frac{p_x \left(4 \pi  T_H+\rho \right){}^2}{8 \pi  \alpha'  m_0 R^2} .
\ee
If the nonrelativistic particle has the momentum $p_i$ at a certain reference time, $t=t_i$, its momentum surpresses exponentially with time
\be
p_x(t)&=&p_i\, \exp \lb -\fr{(4\pi T_H+\r)^2}{8\pi\alpha' R^2 m_0} ( t - t_i)  \rb  .
\ee
In the nonrelativistic limit, temperature and density plays the same role.

\subsection{Microscopic two-point function in the medium \label{sec:3.2}}

Another way to study the medium effect is to look into two-point functions of a local operator. For a CFT, the conformal symmetry restricts the form of a two-point function up to normalization
\be
\bra O(t_1,x_1) \ O( t_2,x_2) \ket \sim \biggl(\fr{1}{ \sqrt{| - (t_1 - t_2 )^2 +(x_1-  x_2)^2|}}\biggr)^{2\D_O} ,
\ee
where $\D_O$ is the conformal dimension of an operator. In medium, interaction with the background matter may break the conformal symmetry and modifies the two-point function. This modification can be understood as the screening effect caused by the medium. In the holographic model, it was argued that a two-point function of the dual QFT can be determined  by  a geodesic length 
\be
\bra O(t_1,x_1) \ O(t_2,x_2) \ket  
\sim e^{- \D_O \, L(t_1, x_1, t_2, x_2) } ,
\ee
where $L(t_1, x_1, t_2, x_2)$ indicates the geodesic length connecting two local operators. For a spatial two-point function with $t_1=t_2$, the corresponding geodesic length is determined by the following holographic formula
\be
L(|x_1- x_2|) =  \int_{x_1}^{ x_2} d x \ \fr{\sqrt{z'^2 + f }} {z \sqrt{f}} .
\ee

The above geodesic length is invariant under the exchange of $x_1$ and $x_2$, so the geodesic has a fixed point at $\bar{x} = (x_1+x_2)/2$ and $z=z_0$. We call this a turning point. The turning point in the $z$-direction represents a maximum value which the geodesic can reach. In other words, the geodesic extends only to $0 \le z \le z_0$. At the turning point, the geodesic must satisfy $dz/dx=0$ to be smooth. Above the geodesic length depends on $x$ implicitly, so there exists a conserved quantity. Using this conserved quantity, we can represent the distance of two operators at the boundary and the geodesic length in the bulk as a function of the turning point 
\be
|x_1- x_2| &=& \int_0^{z_0} dz \fr{2 z}{\sqrt{f (z_0^2 - z^2)}}   ,   \la{Action:HEEl} \\
L(|x_1- x_2|)&=& 2 \int_\e^{z_0} dz \fr{z_0}{z \sqrt{f (z_0^2 - z^2)}}  \la{Action:HEE} ,
\ee
where $\e$ is introduced as a UV cutoff to regularize the UV divergence.

First, we consider a short distance limit corresponding to a UV limit, which on the dual geometry side is realized by taking $z_0 / z_h \ll 1$ and $\r z_0 \ll 1$. Performing the integral \eq{Action:HEEl}  and \eq{Action:HEE} perturbatively leads to the following two-point function 
\be \label{tpscg}
\bra O(t,x_1 )  O(t,x_2) \ket \sim \fr{\exp \ls  -  \fr{  \pi \D_O }{8}  \r  |x_1 - x_2|   -  \fr{2 \pi \D_O }{c} \varepsilon |x_1 - x_2|^2  + \cdots    \rs }{|x_1 - x_2|^{2 \D_O}}  ,   \la{Result:microtwoco}
\ee
where $\varepsilon$ is the energy density of excitations in \eq{Result:exenergy}
\be
\varepsilon \equiv \fr{U_{ex}}{V}=  \fr{\pi   T_H^2}{4 G}  .
\ee
For $\r=\varepsilon=0$, this holographic result reproduces the two-point function expected in a CFT exactly. The above holographic result shows that the two-point function in the medium decreases more rapidly as the density of the matter increases. This is due to the screening effect of the background matters. 


In a long distance limit corresponding to the IR limit, the turning point approaches the horizon and the distance of two local operators diverges. In this long distance limit, the geodesic length can be rewritten as the following form
\be
L(|x_1- x_2|)  = \lim_{z_0 \to z_h} \ls \fr{ |x_1- x_2| }{z_0}+ \fr{2}{ z_0} \int_\e^{z_0} dz \fr{\sqrt{z_0^2 -z^2}}{z \sqrt{f}} \rs.
\ee
In the long distance limit ($|x_1- x_2|\to \infty$), the first term shows a linear divergence but the second term is finite in the IR region. Although a logarithmic divergence in the UV region appears in the second term, it becomes subdominant in the long distance limit. This indicates that the first term gives rise to a leading contribution to the IR two-point function. As a result, the IR two-point function in the quark-gluon medium results in
\be
\bra O(t,x_1)  O(t,x_2) \ket \sim e^{-  |x_1 - x_2| / \xi_c }  , \la{Result:spatialtwopt}
\ee
where the correlation length is identified as
\be \label{colength}
\xi_c= \frac{2}{\D_O \, (4\pi T_H+\r )}.    \la{Result:mIRcorrelation}
\ee
Due to the screening effect by the background medium, the two-point function exponentially suppresses in the IR regime. This suppression becomes fast as the temperature and quark's density increase, as expected. In this case, the correlation length indicates that the spatial two-point function suppresses by a power law in the short-distance limit ($ |x_1 - x_2| < \xi_c$), while it in the long-distance limit ($ |x_1 - x_2|  >  \xi_c$) suppresses exponentially. This result shows that even in the extremal limit ($T_H = 0$), the spatial two-point function exponentially suppresses in the long-distance limit.

This feature together with the temporal two-point function we will study below shows how and why a one-dimensional conformal quantum mechanics occurs in the extremal limit. To see more details, we take into account a temporal two-point function with $x_1=x_2$ which describes the correlation between two time-like operators at the same position \cite{Park:2022mxj}
\be
L (|\ta_1 - \ta_2 | ) = \int_{\ta_1}^{\ta_2}  d \ta \fr{\sqrt{ \dot{z}^2 + f^2}}{z \sqrt{f}}  .
\ee 
From the conservation law, the time interval and geodesic length are determined by the turning point
\be
|\ta_1 - \ta_2 | &=&  \int_0^{z_0} dz \fr{2 \sqrt{f_0} \, z}{f \, \sqrt{f z_0^2 - f_0 z^2}} , 	\la{Relation:timeint}. \\
L (|\ta_1 - \ta_2 | ) &=&  \int_0^{z_0} dz \fr{2  z_0}{ z \, \sqrt{f z_0^2 - f_0 z^2}} ,
\ee
where $f_0$ is the value of $f$ at the turning point. Since the asymptotic geometry of the $p$-brane gas geometry is given by an AdS space, temporal and spatial two-point function in the UV region is given by the CFT one with small correction caused by the matter. From now on, we focus on the IR correlator which can show totally different behavior. 

After rewriting the blackening factor as the form in \eq{Metric:itoRoots}, performing the integral \eq{Relation:timeint} enables us to determine the turning point in terms of the time interval
\begin{align}
z_0 =\frac{2 \, z_{h} z_{in} \, \sin \left(\frac{z_{in}-z_h}{4 z_{in}z_{h}}| \ta_1 - \ta_2 |\right)}{z_{in}-z_{h}+(z_{h}+z_{in}) \, \sin \left(\frac{z_{in}-z_h}{4 z_{in}z_{h}}| \ta_1 - \ta_2 |\right)} .
\end{align}
Using this result, the geodesic length is rewritten as a function of the time interval
\begin{align}
L( | \ta_1 - \ta_2 |) =2 \log \lb \frac{4 z_{in} z_h}{\epsilon \left( z_{in} - z_h\right)} \sin \left(\frac{z_{in} - z_h}{4 z_{in} z_h } | \ta_1 - \ta_2 |\right) \rb .  \la{Result:geolength}
\end{align}
Moving to the Minkowski space via the inverse Wick rotation ($t = - i \ta$), the temporal two-point function is up to a constant multiplication 
\be
\bra O(t_1 ,x)  O( t_2,x) \ket \sim  \ls \fr{(z_{in} - z_h)}{4 z_{in} z_h} \fr{1}{\sinh \left(\frac{z_{in} - z_h}{4 z_{in} z_h } | t_1 - t_2 | \right)} \rs^{2 \D}  .
\ee
In the non-extremal limit with $z_h \ne z_{in}$, the temporal two-point function in the long-time interval limit exponentially suppresses like the previous spatial two-point function 
\be
\lim_{ | t_1 - t_2 |  \to \infty} \bra O(t_1 ,x)  O(t_2,x) \ket \sim  \exp \ls - \frac{(z_{in} - z_h) \D}{2 z_{in} z_h } | t_1 - t_2 |\rs .
\ee
For the BTZ black hole with $\r=0$ (or $z_{in} = - z_h= - 1/\sqrt{M}$), the temporal two-point function in the IR (long-time interval)  limit gives rise to 
\be
\lim_{ | t_1 - t_2 |  \to \infty} \bra O(t_1,x )  O(t_2,x) \ket \sim  e^{- \D \sqrt{M} \, | t_1 - t_2 | } .
\ee 
For the string cloud geometry with $M=0$ ($z_{in} =  \infty$ and $z_h= 1/\r$), on the other hand, it becomes
\be
\lim_{ | t_1 - t_2 |  \to \infty} \bra O(t_1,x )  O(t_2,x) \ket \sim  e^{- \D \r \, | t_1 - t_2 | /2}   .
\ee

In the extremal limit with $z_h = z_{in}$, however, the temporal two-point function shows a different behavior. In the extremal limit, the geodesic length \eq{Result:geolength} reduces to
\be
L( | \ta_1 - \ta_2 |) =2 \log \ls \frac{ | \ta_1 - \ta_2 |  }{\epsilon}  \rs .
\ee
After the inverse Wick rotation, this holographic geodesic length leads to the following temporal two-point function 
\be
\bra O(t_1 ,x)  O(t_2,x) \ket \sim \fr{1}{ | t_1 - t_2 | ^{2 \D}}  ,
\ee
which is equivalent to the CFT's temporal two-point function. As a result, the temporal two-point function in the extremal limit is conformal but the spatial one in \eq{Result:spatialtwopt} is not. This indicates that for the extremal limit of a three-dimensional black hole, a dual two-dimensional CFT in the UV limit flows into a one-dimensional conformal quantum mechanics in the IR region ($| t_1 - t_2 | \gg  1$ and $ |x_1 - x_2| \gg 1$) because the spatial two-point function suppresses exponentially. This is consistent with the fact that the near horizon geometry of an extremal black hole is given by $AdS_2 \times R$ where the spatial direction $R$ decouples.

\section{ Correlation between two macroscopic subsystems in medium \label{sec:4}}

In the previous section, we discussed the microscopic correlation of two local operators in the medium. In this section, we investigate the correlation between macroscopic subsystems instead of local operators. This macroscopic correlation between two subsystmes, $A$ and $B$, is described by the mutual information which gives rise to an upper limit for the collective behavior of microscopic correlations studied before \cite{Wolf:2007tdq,Molina-Vilaplana:2011ydi}
\be
{\rm MI} (A,B) \ge \fr{ \ls  \bra O_A O_B \ket - \bra O_A \ket \bra O_B \ket \rs^2}{2 |O_A|^2 |O_B|^2}  \ge 0  .
\ee
The mutual information is defined in terms of the entanglement entropy studied in Sec. 2
\be \label{mi}
{\rm MI} (A;B)=S_E(A)+S_E(B)-S_E(A\cup B) .
\ee
Due to the subadditivity of entanglement entropy \cite{Casini:2006es,Rangamani:2016dms}, the mutual information must be positive. When the distance of two subsystems is sufficiently far, the mutual information vanishes and there is no quantum correlation. This feature is similar to the correlation length for the microscopic correlation. To study the detail of the mutual information, we take into account two disjoint subsystems, $A$ and $B$ of he same size $l$ and the distance $h$. Then, their positions in the boundary $x$-coordinate are parameterized as
\be
- l - \fr{h}{2} \le A \le  - \fr{h}{2} \quad {\rm and} \quad     \fr{h}{2}  \le B \le  l + \fr{h}{2}  .
\ee
Then, the mutual information is determined by
\begin{align}
{\rm MI} (A,B)  \equiv {\rm MI} (l;h) &= 2S_E (l) - S_E (2l+h) - S_E (h) ,
\end{align}
where $S_E (A \cup B) = S_E (2l+h) + S_E (h)$.

In the UV region with a small subsystem size ($\r l \ll 1$ and $\varepsilon l^2 \ll1$), the entanglement entropy of a subsystem becomes 
\be     \la{Result:2dimHEE}
S_E = \fr{c}{3} \ls  \log \fr{l}{\e}  + \fr{  \pi }{16}  \r  l   +  \fr{\pi}{c} \varepsilon l^2  + \cdots \rs  .
\ee
where the central charge $c = 3 / (2 G)$ represents the degrees of freedom of the dual CFT. This perturbative result in the UV region determines a critical distance $h_c$ where the mutual information vanishes
\be \label{cl}
h_c= l \lb (\sqrt{2}-1)-\frac{(\sqrt{2}-1) \pi \, \r}{16\sqrt{2}} \ l - \frac{2048 \pi \, \varepsilon -  (13-8\sqrt{2}) \pi^2 \, c \r^2 }{1024\sqrt{2} \, c} \ l^2+\cdots  \rb .
\ee
This result shows that the densities of the medium reduces the critical distance. In other words, the correlation of two subsystems vanishes at a shorter distance as the densities $\r$ and $\varepsilon$ are larger. This is because the increase of the medium's densities gives rise to a strong screening effect on the macroscopic correlation which is similar to the microscopic one in \eq{Result:microtwoco}.

Now, let us consider the mutual information in the IR region. Since the mutual information measures the sum of all microscopic correlations, the IR mutual information may lead to a highly nontrivial result involving all UV and IR correlation. However, if we focus on the extreme IR limit of $l \to \infty$, we may expect that only the long-range microscopic correlation dominantly contributes to the mutual information.  To see more details, we rewrite the IR entanglement entropy as the following form
\begin{align}
S_E = \lim_{z_0 \rightarrow z_h} \ls  \frac{R}{4G} \fr{l}{z_0} + \frac{R}{2G z_0} \int_{\epsilon}^{z_0} dz \frac{\sqrt{z_0^2 - z^2}}{z \sqrt{\left(1 - z/z_h \right)\left(1- z/z_{in} \right)}} \rs.
\end{align}
Then, the first term diverges in the IR limit ($l \to \infty$), while the second term is finite up to the UV (logarithmic) divergence. Since the mutual information must be UV finite, the UV divergence of the entanglement entropy does not contribute to the mutual information. The entanglement entropy for $l \to \infty$ leads to
\begin{align}
S_E^{IR} = \frac{1}{4G} \fr{l}{z_h}  + \frac{1}{2}\log\frac{z_h}{\epsilon} + S_C + {\cal O} \ls 1/l \rs ,
\end{align}
with a constant $S_C$ 
\begin{align}
S_C = \frac{1}{G}\left( \sqrt{ \frac{z_{in}}{z_h}} \cot ^{-1}\left(\left(1+\sqrt{2} \sqrt{1-\frac{z_h }{z_{in}}}\right) \sqrt{\frac{z_{in}}{z_h }}\right)- \log \left(\frac{1}{\sqrt{2}}+\frac{1}{2} \sqrt{1-\frac{z_h }{z_{in}}}\right)  \right) ,
\end{align}
where $\log \e$ corresponds to the UV divergence for $\e \to 0$.

Defining a renormalized entanglement entropy as
\be
\bar{S}_E (l)  = S_E (l)  - \frac{1}{2}\log\frac{z_h}{\epsilon} ,
\ee
the mutual information in the IR region reduces to
\begin{align}
{\rm MI} (l;h) &= -\frac{1}{4G} \fr{h}{z_h} - \bar{S}_E  (h) + Sc ,   \la{Result:IRMinfo}
\end{align}
which is UV divergence-free. Introducing a critical distance $h=h_c^{IR} $ where the mutual information vanishes, it is determined as a function of only $z_h/z_{in}$
\be
\fr{h_c^{IR}}{z_h} = F \ls \fr{z_h}{z_{in}}  \rs .
\ee
This result indicates that when the value of $z_h/z_{in}$ is given, the critical distance depends on not the details of $\r$ and $T_H$ but their combination $z_h$.

For the generalized $p$-brane gas geometry, the range of $z_h/z_{in}$ is restricted to be in $-1 \le z_h/z_{in} \le 1$. The generalized $p$-brane gas geometry has two horizons for $0 \le z_h/z_{in} \le 1$. In particular, the extremal limit with $T_H=0$ appears at $z_h/z_{in} = 1$ and the generalized $p$-brane gas geometry at $z_h/z_{in} = 0$ reduces to the string cloud geometry with $M=0$. For $ -1 \le  z_h/z_{in} <0$, on the other hand, it has only one simple root. In this case, the BTZ black hole with $\r=0$ appears at $z_h/z_{in} = -1$. As a consequence, two bounds of the $z_h/z_{in}$ range correspond to the extremal limit at $z_h/z_{in} = 1$ and BTZ black hole at $z_h/z_{in} = -1$.

\begin{figure}
	\centering
	\subfigure[]{\includegraphics[width=0.45\columnwidth]{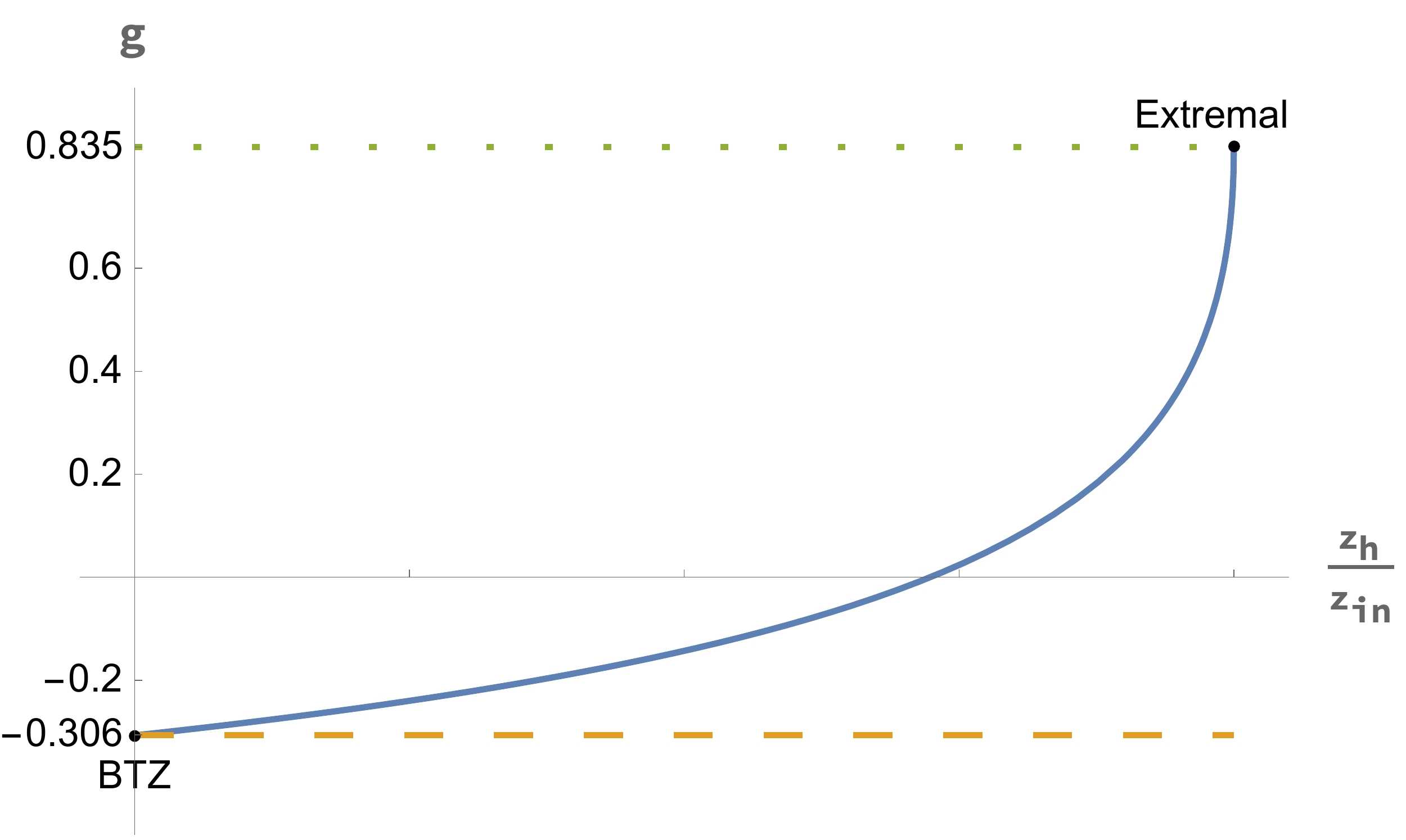}}
	\hspace{0.0em}
	\subfigure[]{\includegraphics[width=0.4\columnwidth]{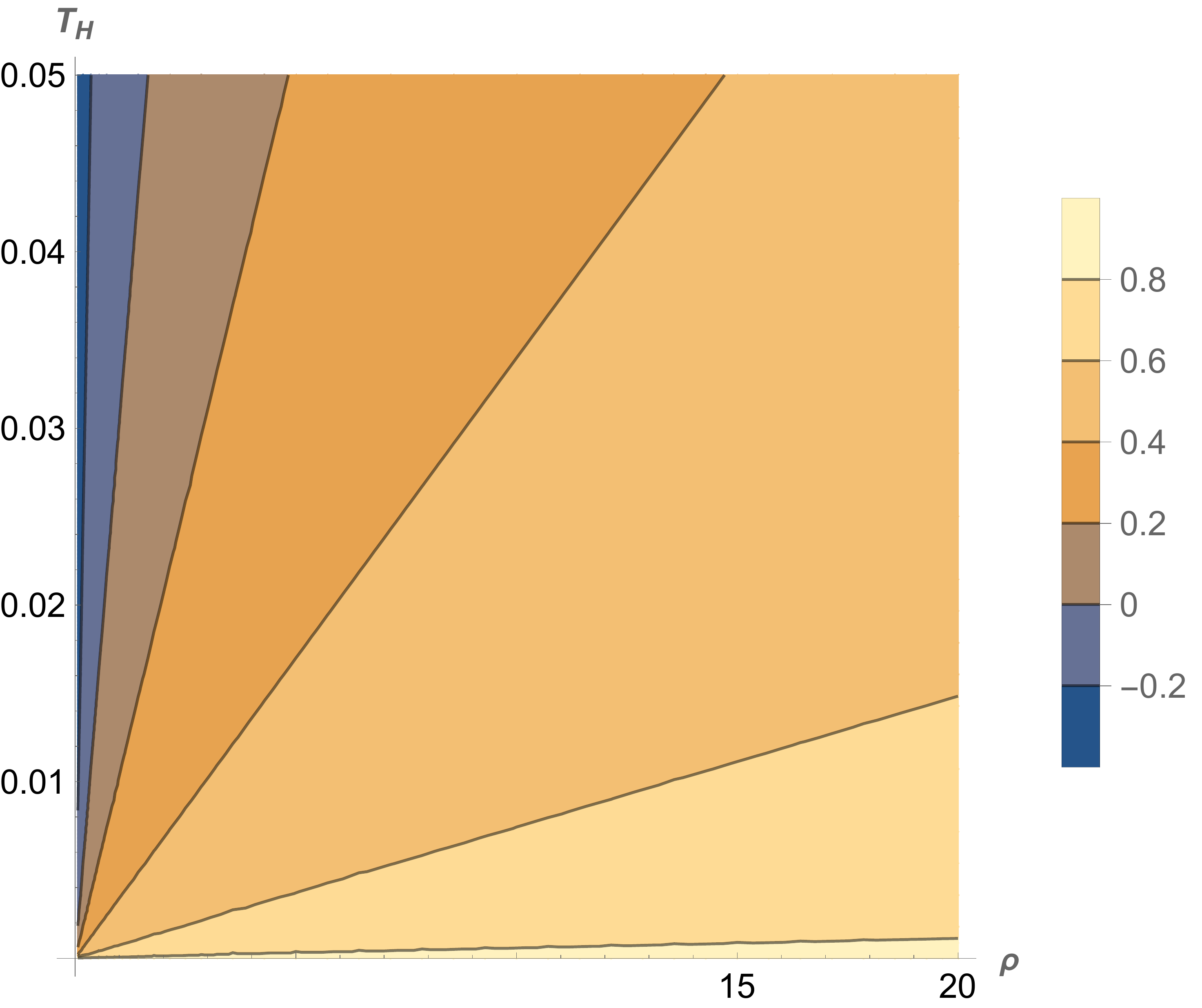}}
	\caption{(a) The value of $g$ (a blue solid curve) where a dotted green and dashed orange line indicate the value of $g$ of the extremal and the BTZ black hole. (b) A contour plot of $g$ depending on $\r$ and $T_H$.}
	\label{ratio}
\end{figure}

For the BTZ black hole with $z_h/z_{in} = -1 $ , the critical distance reduces to
\be
h_c^{IR} = \frac{2  \log 2  }{4 \pi T_H}  .
\ee
Comparing this result with the microscopic correlation length in \eq{Result:mIRcorrelation}, we finally obtain
\be
h_c^{IR} = \fr{ \log 2 }{\D_O} \   \xi_c  .
\ee
This result shows that the critical distance in the IR region is linearly proportional to the correlation length of the IR two-point function. In the extremal limit with $z_h/z_{in} = 1$ and $T_H=0$, on the other hand, we calculate the critical distance numerically and find the relation between $h_c^{IR} $ and $\xi_c $
\be
h_c^{IR}  =  \fr{ 1.835 }{\D_O} \   \xi_c  .
\ee
For a general value of $z_h/z_{in}$, the critical distance in terms of IR correlation length can be rewritten as
\be
h_c^{IR} = \  \lb 1 +  g \ls \fr{z_h}{z_{in}} \rs \rb \  \fr{\xi_c }{\D_O}  .
\ee
In this case, the value of the function $g(z_h/z_{in})$ must be in the following finite range, as shown in Fig. 1, 
 \be
 -0.307 \le g \ls \fr{z_h}{z_{in}} \rs \le 0.835 ,
 \ee
where the lowest limit is given by $\log 2 - 1=  -0.307$. This result indicates that the critical distance of the mutual information behaves similarly to the IR correlation length. Moreover, the macroscopic and microscopic IR correlations in the medium suppresses in the same way, for example, by the inverse of the temperature and density. This is consistent with the screening effect of the background matter.


\section{Discussion \label{sec:5}}

Using the holographic methods, we have investigated microscopic and macroscopic correlations of a medium composed of several matters. To do so, we took into account a generalized $p$-brane gas geometry where bulk $p$-branes correspond to $(p-1)$-dimensional objects on the dual QFT side. Intriguingly, a $p$-brane gas geometry allows a black hole-like geometry with hairs. In the holographic study, it was known that the entanglement temperature for a relativistic QFT is proportional to the inverse of the subsystem size, $T_E \sim l^{-1}$. This feature is universal because it is independent of the dimension and shape of the subsystem. If we further consider a medium including massive objects, we found that the entanglement temperature of a medium shows a different universal feature depending on the dimension of objects, $T_E \sim l^{p-1}$.

In this work, we further considered a three-dimensional generalized $p$-brane gas geometry containing zero- and one-branes which on the dual QFT side correspond to massless fields and massive particles respectively. In this case,  the resulting geometry becomes a black hole with two hairs which are proportional to each $p$-brane's density. Using this geometry dual to a medium consisting of two kinds of matter, we studied the medium effect on the drag force and correlation function of a local operator. In the short distance and time interval limit which corresponds to the UV limit, the spatial and temporal two-point functions become the CFT's one with small corrections, because the medium effect in the UV limit becomes subdominant. In the long distance and time interval corresponding to the IR limit, the two-point functions suppress exponentially due to the screening effect of the background matter. We showed that the correlation length is inversely proportional to the temperature and density. 

The medium composed of two kinds of matter is the dual of a generalized $p$-brane gas geometry with two kinds of $p$-brane which resembles a black hole having two hairs. We showed that this black hole geometry allows an extremal limit (or zero temperature limit). The asymptote of this geometry is given by AdS$_3$, whereas the near horizon geometry in the extremal limit reduces to AdS$_2 \times R$. These geometries in the UV and IR limits indicate that the dual QFT flows from a two-dimensional UV CFT into a one-dimensional conformal quantum mechanics where the spatial direction decouples. Using the geodesic extending to the generalized $p$-brane gas geometry, we studied how the temporal and spatial two-point functions evolve along the RG flow. We showed that the spatial two-point function always suppresses exponentially in the IR (or long distance) limit. However, we found that the temporal two-point function in the extremal limit is conformal. This fact indicates that the spatial direction decouples in the IR limit due to the exponential suppression of the two-point function. As a result, the IR theory becomes a one-dimensional conformal quantum mechanics which is consistent with the near horizon geometry discussed before.

Lastly, we also investigated the mutual information representing the correlation between two macroscopic subsystems. It was known that, as the distance between two subsystems becomes large, the quantum entanglement between two subsystems gradually decreases and then finally vanishes at a certain critical distance. A similar feature also appears in the two-point function of a local operator. In the IR limit, if the distance of two local operators is shorter than the correlation length, the spatial two-point function decreases by a power law as the distance becomes long. However, if the distance is longer than the correlation length, the two-point function suppresses exponentially. Due to these similar features, we compared the mutual information with the two-point function in the IR limit. The mutual information may be regarded as the collective behavior of microscopic correlations. Therefore, we expected that the IR mutual information and two-point function lead to a similar feature. By applying the holographic methods, we explicitly showed that the critical distance of the mutual information behaves in a similar way to the spatial two-point function in the IR region.

\vspace{1cm}


{\bf Acknowledgement}

C. Park was supported by the National Research Foundation of Korea(NRF) grant funded by the Korea government(MSIT) (No. NRF-2019R1A2C1006639).  J. H. Lee was supported by the National Research Foundation of Korea(NRF) grant funded by the Korea government(MSIT) (No. NRF-2021R1C1C2008737). S-J. Kim was supported by Basic Science Research Program through the National Research Foundation of Korea, funded by the Ministry of Education grant (No. NRF-2021R1I1A1A01052821)



	\bibliographystyle{apsrev4-1}

\bibliography{ref.bib}

\begin{thebibliography}{50}%
\makeatletter
\providecommand \@ifxundefined [1]{%
 \@ifx{#1\undefined}
}%
\providecommand \@ifnum [1]{%
 \ifnum #1\expandafter \@firstoftwo
 \else \expandafter \@secondoftwo
 \fi
}%
\providecommand \@ifx [1]{%
 \ifx #1\expandafter \@firstoftwo
 \else \expandafter \@secondoftwo
 \fi
}%
\providecommand \natexlab [1]{#1}%
\providecommand \enquote  [1]{``#1''}%
\providecommand \bibnamefont  [1]{#1}%
\providecommand \bibfnamefont [1]{#1}%
\providecommand \citenamefont [1]{#1}%
\providecommand \href@noop [0]{\@secondoftwo}%
\providecommand \href [0]{\begingroup \@sanitize@url \@href}%
\providecommand \@href[1]{\@@startlink{#1}\@@href}%
\providecommand \@@href[1]{\endgroup#1\@@endlink}%
\providecommand \@sanitize@url [0]{\catcode `\\12\catcode `\$12\catcode
  `\&12\catcode `\#12\catcode `\^12\catcode `\_12\catcode `\%12\relax}%
\providecommand \@@startlink[1]{}%
\providecommand \@@endlink[0]{}%
\providecommand \url  [0]{\begingroup\@sanitize@url \@url }%
\providecommand \@url [1]{\endgroup\@href {#1}{\urlprefix }}%
\providecommand \urlprefix  [0]{URL }%
\providecommand \Eprint [0]{\href }%
\providecommand \doibase [0]{http://dx.doi.org/}%
\providecommand \selectlanguage [0]{\@gobble}%
\providecommand \bibinfo  [0]{\@secondoftwo}%
\providecommand \bibfield  [0]{\@secondoftwo}%
\providecommand \translation [1]{[#1]}%
\providecommand \BibitemOpen [0]{}%
\providecommand \bibitemStop [0]{}%
\providecommand \bibitemNoStop [0]{.\EOS\space}%
\providecommand \EOS [0]{\spacefactor3000\relax}%
\providecommand \BibitemShut  [1]{\csname bibitem#1\endcsname}%
\let\auto@bib@innerbib\@empty
\bibitem [{\citenamefont {Maldacena}(1999)}]{Maldacena:1997re}%
  \BibitemOpen
  \bibfield  {author} {\bibinfo {author} {\bibfnamefont {J.~M.}\ \bibnamefont
  {Maldacena}},\ }\href {\doibase 10.1023/A:1026654312961,
  10.4310/ATMP.1998.v2.n2.a1} {\bibfield  {journal} {\bibinfo  {journal} {Int.
  J. Theor. Phys.}\ }\textbf {\bibinfo {volume} {38}},\ \bibinfo {pages} {1113}
  (\bibinfo {year} {1999})},\ \bibinfo {note} {[Adv. Theor. Math.
  Phys.2,231(1998)]},\ \Eprint {http://arxiv.org/abs/hep-th/9711200}
  {arXiv:hep-th/9711200 [hep-th]} \BibitemShut {NoStop}%
\bibitem [{\citenamefont {Witten}(1998)}]{Witten:1998qj}%
  \BibitemOpen
  \bibfield  {author} {\bibinfo {author} {\bibfnamefont {E.}~\bibnamefont
  {Witten}},\ }\href {\doibase 10.4310/ATMP.1998.v2.n2.a2} {\bibfield
  {journal} {\bibinfo  {journal} {Adv. Theor. Math. Phys.}\ }\textbf {\bibinfo
  {volume} {2}},\ \bibinfo {pages} {253} (\bibinfo {year} {1998})},\ \Eprint
  {http://arxiv.org/abs/hep-th/9802150} {arXiv:hep-th/9802150 [hep-th]}
  \BibitemShut {NoStop}%
\bibitem [{\citenamefont {Gubser}\ \emph {et~al.}(1998)\citenamefont {Gubser},
  \citenamefont {Klebanov},\ and\ \citenamefont {Polyakov}}]{Gubser:1998bc}%
  \BibitemOpen
  \bibfield  {author} {\bibinfo {author} {\bibfnamefont {S.~S.}\ \bibnamefont
  {Gubser}}, \bibinfo {author} {\bibfnamefont {I.~R.}\ \bibnamefont
  {Klebanov}}, \ and\ \bibinfo {author} {\bibfnamefont {A.~M.}\ \bibnamefont
  {Polyakov}},\ }\href {\doibase 10.1016/S0370-2693(98)00377-3} {\bibfield
  {journal} {\bibinfo  {journal} {Phys. Lett. B}\ }\textbf {\bibinfo {volume}
  {428}},\ \bibinfo {pages} {105} (\bibinfo {year} {1998})},\ \Eprint
  {http://arxiv.org/abs/hep-th/9802109} {arXiv:hep-th/9802109} \BibitemShut
  {NoStop}%
\bibitem [{\citenamefont {Maldacena}(2002)}]{Maldacena:2002mn}%
  \BibitemOpen
  \bibfield  {author} {\bibinfo {author} {\bibfnamefont {J.~M.}\ \bibnamefont
  {Maldacena}},\ }in\ \href {\doibase 10.1007/0-387-24992-3_3} {\emph {\bibinfo
  {booktitle} {{School on Quantum Gravity}}}}\ (\bibinfo {year} {2002})\ pp.\
  \bibinfo {pages} {91--150}\BibitemShut {NoStop}%
\bibitem [{\citenamefont {Hartnoll}\ \emph
  {et~al.}(2008{\natexlab{a}})\citenamefont {Hartnoll}, \citenamefont
  {Herzog},\ and\ \citenamefont {Horowitz}}]{Hartnoll:2008kx}%
  \BibitemOpen
  \bibfield  {author} {\bibinfo {author} {\bibfnamefont {S.~A.}\ \bibnamefont
  {Hartnoll}}, \bibinfo {author} {\bibfnamefont {C.~P.}\ \bibnamefont
  {Herzog}}, \ and\ \bibinfo {author} {\bibfnamefont {G.~T.}\ \bibnamefont
  {Horowitz}},\ }\href {\doibase 10.1088/1126-6708/2008/12/015} {\bibfield
  {journal} {\bibinfo  {journal} {JHEP}\ }\textbf {\bibinfo {volume} {12}},\
  \bibinfo {pages} {015} (\bibinfo {year} {2008}{\natexlab{a}})},\ \Eprint
  {http://arxiv.org/abs/0810.1563} {arXiv:0810.1563 [hep-th]} \BibitemShut
  {NoStop}%
\bibitem [{\citenamefont {Maldacena}\ and\ \citenamefont
  {Pimentel}(2013)}]{Maldacena:2012xp}%
  \BibitemOpen
  \bibfield  {author} {\bibinfo {author} {\bibfnamefont {J.}~\bibnamefont
  {Maldacena}}\ and\ \bibinfo {author} {\bibfnamefont {G.~L.}\ \bibnamefont
  {Pimentel}},\ }\href {\doibase 10.1007/JHEP02(2013)038} {\bibfield  {journal}
  {\bibinfo  {journal} {JHEP}\ }\textbf {\bibinfo {volume} {02}},\ \bibinfo
  {pages} {038} (\bibinfo {year} {2013})},\ \Eprint
  {http://arxiv.org/abs/1210.7244} {arXiv:1210.7244 [hep-th]} \BibitemShut
  {NoStop}%
\bibitem [{\citenamefont {Myers}\ and\ \citenamefont
  {Singh}(2012)}]{Myers:2012ed}%
  \BibitemOpen
  \bibfield  {author} {\bibinfo {author} {\bibfnamefont {R.~C.}\ \bibnamefont
  {Myers}}\ and\ \bibinfo {author} {\bibfnamefont {A.}~\bibnamefont {Singh}},\
  }\href {\doibase 10.1007/JHEP04(2012)122} {\bibfield  {journal} {\bibinfo
  {journal} {JHEP}\ }\textbf {\bibinfo {volume} {04}},\ \bibinfo {pages} {122}
  (\bibinfo {year} {2012})},\ \Eprint {http://arxiv.org/abs/1202.2068}
  {arXiv:1202.2068 [hep-th]} \BibitemShut {NoStop}%
\bibitem [{\citenamefont {Bueno}\ \emph {et~al.}(2015)\citenamefont {Bueno},
  \citenamefont {Myers},\ and\ \citenamefont {Witczak-Krempa}}]{Bueno:2015rda}%
  \BibitemOpen
  \bibfield  {author} {\bibinfo {author} {\bibfnamefont {P.}~\bibnamefont
  {Bueno}}, \bibinfo {author} {\bibfnamefont {R.~C.}\ \bibnamefont {Myers}}, \
  and\ \bibinfo {author} {\bibfnamefont {W.}~\bibnamefont {Witczak-Krempa}},\
  }\href {\doibase 10.1103/PhysRevLett.115.021602} {\bibfield  {journal}
  {\bibinfo  {journal} {Phys. Rev. Lett.}\ }\textbf {\bibinfo {volume} {115}},\
  \bibinfo {pages} {021602} (\bibinfo {year} {2015})},\ \Eprint
  {http://arxiv.org/abs/1505.04804} {arXiv:1505.04804 [hep-th]} \BibitemShut
  {NoStop}%
\bibitem [{\citenamefont {Kim}\ \emph {et~al.}(2019)\citenamefont {Kim},
  \citenamefont {Park}, \citenamefont {Hun~Lee},\ and\ \citenamefont
  {Ahn}}]{Kim:2018mgz}%
  \BibitemOpen
  \bibfield  {author} {\bibinfo {author} {\bibfnamefont {K.~K.}\ \bibnamefont
  {Kim}}, \bibinfo {author} {\bibfnamefont {C.}~\bibnamefont {Park}}, \bibinfo
  {author} {\bibfnamefont {J.}~\bibnamefont {Hun~Lee}}, \ and\ \bibinfo
  {author} {\bibfnamefont {B.}~\bibnamefont {Ahn}},\ }\href {\doibase
  10.1140/epjc/s10052-019-6888-z} {\bibfield  {journal} {\bibinfo  {journal}
  {Eur. Phys. J. C}\ }\textbf {\bibinfo {volume} {79}},\ \bibinfo {pages} {377}
  (\bibinfo {year} {2019})},\ \Eprint {http://arxiv.org/abs/1804.00412}
  {arXiv:1804.00412 [hep-th]} \BibitemShut {NoStop}%
\bibitem [{\citenamefont {Koh}\ \emph {et~al.}(2018)\citenamefont {Koh},
  \citenamefont {Hun~Lee}, \citenamefont {Park},\ and\ \citenamefont
  {Ro}}]{Koh:2018rsw}%
  \BibitemOpen
  \bibfield  {author} {\bibinfo {author} {\bibfnamefont {S.}~\bibnamefont
  {Koh}}, \bibinfo {author} {\bibfnamefont {J.}~\bibnamefont {Hun~Lee}},
  \bibinfo {author} {\bibfnamefont {C.}~\bibnamefont {Park}}, \ and\ \bibinfo
  {author} {\bibfnamefont {D.}~\bibnamefont {Ro}},\ }\href@noop {} {\
  (\bibinfo {year} {2018})},\ \Eprint {http://arxiv.org/abs/1806.01092}
  {arXiv:1806.01092 [hep-th]} \BibitemShut {NoStop}%
\bibitem [{\citenamefont {Park}\ \emph {et~al.}(2018)\citenamefont {Park},
  \citenamefont {Ro},\ and\ \citenamefont {Hun~Lee}}]{Park:2018ebm}%
  \BibitemOpen
  \bibfield  {author} {\bibinfo {author} {\bibfnamefont {C.}~\bibnamefont
  {Park}}, \bibinfo {author} {\bibfnamefont {D.}~\bibnamefont {Ro}}, \ and\
  \bibinfo {author} {\bibfnamefont {J.}~\bibnamefont {Hun~Lee}},\ }\href
  {\doibase 10.1007/JHEP11(2018)165} {\bibfield  {journal} {\bibinfo  {journal}
  {JHEP}\ }\textbf {\bibinfo {volume} {11}},\ \bibinfo {pages} {165} (\bibinfo
  {year} {2018})},\ \Eprint {http://arxiv.org/abs/1806.09072} {arXiv:1806.09072
  [hep-th]} \BibitemShut {NoStop}%
\bibitem [{\citenamefont {Cooper}\ \emph {et~al.}(2019)\citenamefont {Cooper},
  \citenamefont {Rozali}, \citenamefont {Swingle}, \citenamefont
  {Van~Raamsdonk}, \citenamefont {Waddell},\ and\ \citenamefont
  {Wakeham}}]{Cooper:2018cmb}%
  \BibitemOpen
  \bibfield  {author} {\bibinfo {author} {\bibfnamefont {S.}~\bibnamefont
  {Cooper}}, \bibinfo {author} {\bibfnamefont {M.}~\bibnamefont {Rozali}},
  \bibinfo {author} {\bibfnamefont {B.}~\bibnamefont {Swingle}}, \bibinfo
  {author} {\bibfnamefont {M.}~\bibnamefont {Van~Raamsdonk}}, \bibinfo {author}
  {\bibfnamefont {C.}~\bibnamefont {Waddell}}, \ and\ \bibinfo {author}
  {\bibfnamefont {D.}~\bibnamefont {Wakeham}},\ }\href {\doibase
  10.1007/JHEP07(2019)065} {\bibfield  {journal} {\bibinfo  {journal} {JHEP}\
  }\textbf {\bibinfo {volume} {07}},\ \bibinfo {pages} {065} (\bibinfo {year}
  {2019})},\ \Eprint {http://arxiv.org/abs/1810.10601} {arXiv:1810.10601
  [hep-th]} \BibitemShut {NoStop}%
\bibitem [{\citenamefont {Almheiri}\ \emph {et~al.}(2020)\citenamefont
  {Almheiri}, \citenamefont {Mahajan}, \citenamefont {Maldacena},\ and\
  \citenamefont {Zhao}}]{Almheiri:2019hni}%
  \BibitemOpen
  \bibfield  {author} {\bibinfo {author} {\bibfnamefont {A.}~\bibnamefont
  {Almheiri}}, \bibinfo {author} {\bibfnamefont {R.}~\bibnamefont {Mahajan}},
  \bibinfo {author} {\bibfnamefont {J.}~\bibnamefont {Maldacena}}, \ and\
  \bibinfo {author} {\bibfnamefont {Y.}~\bibnamefont {Zhao}},\ }\href {\doibase
  10.1007/JHEP03(2020)149} {\bibfield  {journal} {\bibinfo  {journal} {JHEP}\
  }\textbf {\bibinfo {volume} {03}},\ \bibinfo {pages} {149} (\bibinfo {year}
  {2020})},\ \Eprint {http://arxiv.org/abs/1908.10996} {arXiv:1908.10996
  [hep-th]} \BibitemShut {NoStop}%
\bibitem [{\citenamefont {Policastro}\ \emph {et~al.}(2001)\citenamefont
  {Policastro}, \citenamefont {Son},\ and\ \citenamefont
  {Starinets}}]{Policastro:2001yc}%
  \BibitemOpen
  \bibfield  {author} {\bibinfo {author} {\bibfnamefont {G.}~\bibnamefont
  {Policastro}}, \bibinfo {author} {\bibfnamefont {D.~T.}\ \bibnamefont {Son}},
  \ and\ \bibinfo {author} {\bibfnamefont {A.~O.}\ \bibnamefont {Starinets}},\
  }\href {\doibase 10.1103/PhysRevLett.87.081601} {\bibfield  {journal}
  {\bibinfo  {journal} {Phys. Rev. Lett.}\ }\textbf {\bibinfo {volume} {87}},\
  \bibinfo {pages} {081601} (\bibinfo {year} {2001})},\ \Eprint
  {http://arxiv.org/abs/hep-th/0104066} {arXiv:hep-th/0104066} \BibitemShut
  {NoStop}%
\bibitem [{\citenamefont {Hartnoll}\ \emph
  {et~al.}(2008{\natexlab{b}})\citenamefont {Hartnoll}, \citenamefont
  {Herzog},\ and\ \citenamefont {Horowitz}}]{Hartnoll:2008vx}%
  \BibitemOpen
  \bibfield  {author} {\bibinfo {author} {\bibfnamefont {S.~A.}\ \bibnamefont
  {Hartnoll}}, \bibinfo {author} {\bibfnamefont {C.~P.}\ \bibnamefont
  {Herzog}}, \ and\ \bibinfo {author} {\bibfnamefont {G.~T.}\ \bibnamefont
  {Horowitz}},\ }\href {\doibase 10.1103/PhysRevLett.101.031601} {\bibfield
  {journal} {\bibinfo  {journal} {Phys. Rev. Lett.}\ }\textbf {\bibinfo
  {volume} {101}},\ \bibinfo {pages} {031601} (\bibinfo {year}
  {2008}{\natexlab{b}})},\ \Eprint {http://arxiv.org/abs/0803.3295}
  {arXiv:0803.3295 [hep-th]} \BibitemShut {NoStop}%
\bibitem [{\citenamefont {Swingle}(2012)}]{Swingle:2009bg}%
  \BibitemOpen
  \bibfield  {author} {\bibinfo {author} {\bibfnamefont {B.}~\bibnamefont
  {Swingle}},\ }\href {\doibase 10.1103/PhysRevD.86.065007} {\bibfield
  {journal} {\bibinfo  {journal} {Phys. Rev. D}\ }\textbf {\bibinfo {volume}
  {86}},\ \bibinfo {pages} {065007} (\bibinfo {year} {2012})},\ \Eprint
  {http://arxiv.org/abs/0905.1317} {arXiv:0905.1317 [cond-mat.str-el]}
  \BibitemShut {NoStop}%
\bibitem [{\citenamefont {Karch}\ \emph {et~al.}(2010)\citenamefont {Karch},
  \citenamefont {Maciejko},\ and\ \citenamefont {Takayanagi}}]{Karch:2010mn}%
  \BibitemOpen
  \bibfield  {author} {\bibinfo {author} {\bibfnamefont {A.}~\bibnamefont
  {Karch}}, \bibinfo {author} {\bibfnamefont {J.}~\bibnamefont {Maciejko}}, \
  and\ \bibinfo {author} {\bibfnamefont {T.}~\bibnamefont {Takayanagi}},\
  }\href {\doibase 10.1103/PhysRevD.82.126003} {\bibfield  {journal} {\bibinfo
  {journal} {Phys. Rev. D}\ }\textbf {\bibinfo {volume} {82}},\ \bibinfo
  {pages} {126003} (\bibinfo {year} {2010})},\ \Eprint
  {http://arxiv.org/abs/1009.2991} {arXiv:1009.2991 [hep-th]} \BibitemShut
  {NoStop}%
\bibitem [{\citenamefont {Nozaki}\ \emph {et~al.}(2012)\citenamefont {Nozaki},
  \citenamefont {Ryu},\ and\ \citenamefont {Takayanagi}}]{Nozaki:2012zj}%
  \BibitemOpen
  \bibfield  {author} {\bibinfo {author} {\bibfnamefont {M.}~\bibnamefont
  {Nozaki}}, \bibinfo {author} {\bibfnamefont {S.}~\bibnamefont {Ryu}}, \ and\
  \bibinfo {author} {\bibfnamefont {T.}~\bibnamefont {Takayanagi}},\ }\href
  {\doibase 10.1007/JHEP10(2012)193} {\bibfield  {journal} {\bibinfo  {journal}
  {JHEP}\ }\textbf {\bibinfo {volume} {10}},\ \bibinfo {pages} {193} (\bibinfo
  {year} {2012})},\ \Eprint {http://arxiv.org/abs/1208.3469} {arXiv:1208.3469
  [hep-th]} \BibitemShut {NoStop}%
\bibitem [{\citenamefont {Erdmenger}\ \emph {et~al.}(2016)\citenamefont
  {Erdmenger}, \citenamefont {Flory}, \citenamefont {Hoyos}, \citenamefont
  {Newrzella},\ and\ \citenamefont {Wu}}]{Erdmenger:2015spo}%
  \BibitemOpen
  \bibfield  {author} {\bibinfo {author} {\bibfnamefont {J.}~\bibnamefont
  {Erdmenger}}, \bibinfo {author} {\bibfnamefont {M.}~\bibnamefont {Flory}},
  \bibinfo {author} {\bibfnamefont {C.}~\bibnamefont {Hoyos}}, \bibinfo
  {author} {\bibfnamefont {M.-N.}\ \bibnamefont {Newrzella}}, \ and\ \bibinfo
  {author} {\bibfnamefont {J.~M.~S.}\ \bibnamefont {Wu}},\ }\href {\doibase
  10.1002/prop.201500099} {\bibfield  {journal} {\bibinfo  {journal} {Fortsch.
  Phys.}\ }\textbf {\bibinfo {volume} {64}},\ \bibinfo {pages} {109} (\bibinfo
  {year} {2016})},\ \Eprint {http://arxiv.org/abs/1511.03666} {arXiv:1511.03666
  [hep-th]} \BibitemShut {NoStop}%
\bibitem [{\citenamefont {Lashkari}\ and\ \citenamefont
  {Van~Raamsdonk}(2016)}]{Lashkari:2015hha}%
  \BibitemOpen
  \bibfield  {author} {\bibinfo {author} {\bibfnamefont {N.}~\bibnamefont
  {Lashkari}}\ and\ \bibinfo {author} {\bibfnamefont {M.}~\bibnamefont
  {Van~Raamsdonk}},\ }\href {\doibase 10.1007/JHEP04(2016)153} {\bibfield
  {journal} {\bibinfo  {journal} {JHEP}\ }\textbf {\bibinfo {volume} {04}},\
  \bibinfo {pages} {153} (\bibinfo {year} {2016})},\ \Eprint
  {http://arxiv.org/abs/1508.00897} {arXiv:1508.00897 [hep-th]} \BibitemShut
  {NoStop}%
\bibitem [{\citenamefont {Park}\ and\ \citenamefont
  {Hun~Lee}(2018)}]{Park:2017ray}%
  \BibitemOpen
  \bibfield  {author} {\bibinfo {author} {\bibfnamefont {C.}~\bibnamefont
  {Park}}\ and\ \bibinfo {author} {\bibfnamefont {J.}~\bibnamefont {Hun~Lee}},\
  }\href {\doibase 10.1142/S0217751X18500161} {\bibfield  {journal} {\bibinfo
  {journal} {Int. J. Mod. Phys. A}\ }\textbf {\bibinfo {volume} {33}},\
  \bibinfo {pages} {1850016} (\bibinfo {year} {2018})},\ \Eprint
  {http://arxiv.org/abs/1707.05482} {arXiv:1707.05482 [hep-th]} \BibitemShut
  {NoStop}%
\bibitem [{\citenamefont {Akutagawa}\ \emph {et~al.}(2020)\citenamefont
  {Akutagawa}, \citenamefont {Hashimoto},\ and\ \citenamefont
  {Sumimoto}}]{Akutagawa:2020yeo}%
  \BibitemOpen
  \bibfield  {author} {\bibinfo {author} {\bibfnamefont {T.}~\bibnamefont
  {Akutagawa}}, \bibinfo {author} {\bibfnamefont {K.}~\bibnamefont
  {Hashimoto}}, \ and\ \bibinfo {author} {\bibfnamefont {T.}~\bibnamefont
  {Sumimoto}},\ }\href {\doibase 10.1103/PhysRevD.102.026020} {\bibfield
  {journal} {\bibinfo  {journal} {Phys. Rev. D}\ }\textbf {\bibinfo {volume}
  {102}},\ \bibinfo {pages} {026020} (\bibinfo {year} {2020})},\ \Eprint
  {http://arxiv.org/abs/2005.02636} {arXiv:2005.02636 [hep-th]} \BibitemShut
  {NoStop}%
\bibitem [{\citenamefont {Shenker}\ and\ \citenamefont
  {Stanford}(2014)}]{Shenker:2013pqa}%
  \BibitemOpen
  \bibfield  {author} {\bibinfo {author} {\bibfnamefont {S.~H.}\ \bibnamefont
  {Shenker}}\ and\ \bibinfo {author} {\bibfnamefont {D.}~\bibnamefont
  {Stanford}},\ }\href {\doibase 10.1007/JHEP03(2014)067} {\bibfield  {journal}
  {\bibinfo  {journal} {JHEP}\ }\textbf {\bibinfo {volume} {03}},\ \bibinfo
  {pages} {067} (\bibinfo {year} {2014})},\ \Eprint
  {http://arxiv.org/abs/1306.0622} {arXiv:1306.0622 [hep-th]} \BibitemShut
  {NoStop}%
\bibitem [{\citenamefont {Maldacena}(1998)}]{Maldacena:1998im}%
  \BibitemOpen
  \bibfield  {author} {\bibinfo {author} {\bibfnamefont {J.~M.}\ \bibnamefont
  {Maldacena}},\ }\href {\doibase 10.1103/PhysRevLett.80.4859} {\bibfield
  {journal} {\bibinfo  {journal} {Phys. Rev. Lett.}\ }\textbf {\bibinfo
  {volume} {80}},\ \bibinfo {pages} {4859} (\bibinfo {year} {1998})},\ \Eprint
  {http://arxiv.org/abs/hep-th/9803002} {arXiv:hep-th/9803002} \BibitemShut
  {NoStop}%
\bibitem [{\citenamefont {Ryu}\ and\ \citenamefont
  {Takayanagi}(2006{\natexlab{a}})}]{Ryu:2006bv}%
  \BibitemOpen
  \bibfield  {author} {\bibinfo {author} {\bibfnamefont {S.}~\bibnamefont
  {Ryu}}\ and\ \bibinfo {author} {\bibfnamefont {T.}~\bibnamefont
  {Takayanagi}},\ }\href {\doibase 10.1103/PhysRevLett.96.181602} {\bibfield
  {journal} {\bibinfo  {journal} {Phys. Rev. Lett.}\ }\textbf {\bibinfo
  {volume} {96}},\ \bibinfo {pages} {181602} (\bibinfo {year}
  {2006}{\natexlab{a}})},\ \Eprint {http://arxiv.org/abs/hep-th/0603001}
  {arXiv:hep-th/0603001 [hep-th]} \BibitemShut {NoStop}%
\bibitem [{\citenamefont {Ryu}\ and\ \citenamefont
  {Takayanagi}(2006{\natexlab{b}})}]{Ryu:2006ef}%
  \BibitemOpen
  \bibfield  {author} {\bibinfo {author} {\bibfnamefont {S.}~\bibnamefont
  {Ryu}}\ and\ \bibinfo {author} {\bibfnamefont {T.}~\bibnamefont
  {Takayanagi}},\ }\href {\doibase 10.1088/1126-6708/2006/08/045} {\bibfield
  {journal} {\bibinfo  {journal} {JHEP}\ }\textbf {\bibinfo {volume} {08}},\
  \bibinfo {pages} {045} (\bibinfo {year} {2006}{\natexlab{b}})},\ \Eprint
  {http://arxiv.org/abs/hep-th/0605073} {arXiv:hep-th/0605073 [hep-th]}
  \BibitemShut {NoStop}%
\bibitem [{\citenamefont {Hubeny}\ \emph {et~al.}(2007)\citenamefont {Hubeny},
  \citenamefont {Rangamani},\ and\ \citenamefont {Takayanagi}}]{Hubeny:2007xt}%
  \BibitemOpen
  \bibfield  {author} {\bibinfo {author} {\bibfnamefont {V.~E.}\ \bibnamefont
  {Hubeny}}, \bibinfo {author} {\bibfnamefont {M.}~\bibnamefont {Rangamani}}, \
  and\ \bibinfo {author} {\bibfnamefont {T.}~\bibnamefont {Takayanagi}},\
  }\href {\doibase 10.1088/1126-6708/2007/07/062} {\bibfield  {journal}
  {\bibinfo  {journal} {JHEP}\ }\textbf {\bibinfo {volume} {07}},\ \bibinfo
  {pages} {062} (\bibinfo {year} {2007})},\ \Eprint
  {http://arxiv.org/abs/0705.0016} {arXiv:0705.0016 [hep-th]} \BibitemShut
  {NoStop}%
\bibitem [{\citenamefont {Casini}\ \emph {et~al.}(2011)\citenamefont {Casini},
  \citenamefont {Huerta},\ and\ \citenamefont {Myers}}]{Casini:2011kv}%
  \BibitemOpen
  \bibfield  {author} {\bibinfo {author} {\bibfnamefont {H.}~\bibnamefont
  {Casini}}, \bibinfo {author} {\bibfnamefont {M.}~\bibnamefont {Huerta}}, \
  and\ \bibinfo {author} {\bibfnamefont {R.~C.}\ \bibnamefont {Myers}},\ }\href
  {\doibase 10.1007/JHEP05(2011)036} {\bibfield  {journal} {\bibinfo  {journal}
  {JHEP}\ }\textbf {\bibinfo {volume} {05}},\ \bibinfo {pages} {036} (\bibinfo
  {year} {2011})},\ \Eprint {http://arxiv.org/abs/1102.0440} {arXiv:1102.0440
  [hep-th]} \BibitemShut {NoStop}%
\bibitem [{\citenamefont {Nishioka}\ \emph {et~al.}(2009)\citenamefont
  {Nishioka}, \citenamefont {Ryu},\ and\ \citenamefont
  {Takayanagi}}]{Nishioka:2009un}%
  \BibitemOpen
  \bibfield  {author} {\bibinfo {author} {\bibfnamefont {T.}~\bibnamefont
  {Nishioka}}, \bibinfo {author} {\bibfnamefont {S.}~\bibnamefont {Ryu}}, \
  and\ \bibinfo {author} {\bibfnamefont {T.}~\bibnamefont {Takayanagi}},\
  }\href {\doibase 10.1088/1751-8113/42/50/504008} {\bibfield  {journal}
  {\bibinfo  {journal} {J. Phys. A}\ }\textbf {\bibinfo {volume} {42}},\
  \bibinfo {pages} {504008} (\bibinfo {year} {2009})},\ \Eprint
  {http://arxiv.org/abs/0905.0932} {arXiv:0905.0932 [hep-th]} \BibitemShut
  {NoStop}%
\bibitem [{\citenamefont {Takayanagi}(2012)}]{Takayanagi:2012kg}%
  \BibitemOpen
  \bibfield  {author} {\bibinfo {author} {\bibfnamefont {T.}~\bibnamefont
  {Takayanagi}},\ }\href {\doibase 10.1088/0264-9381/29/15/153001} {\bibfield
  {journal} {\bibinfo  {journal} {Class. Quant. Grav.}\ }\textbf {\bibinfo
  {volume} {29}},\ \bibinfo {pages} {153001} (\bibinfo {year} {2012})},\
  \Eprint {http://arxiv.org/abs/1204.2450} {arXiv:1204.2450 [gr-qc]}
  \BibitemShut {NoStop}%
\bibitem [{\citenamefont {Kim}\ \emph {et~al.}(2012)\citenamefont {Kim},
  \citenamefont {Kim},\ and\ \citenamefont {Hun~Lee}}]{Kim:2012tu}%
  \BibitemOpen
  \bibfield  {author} {\bibinfo {author} {\bibfnamefont {H.}~\bibnamefont
  {Kim}}, \bibinfo {author} {\bibfnamefont {N.}~\bibnamefont {Kim}}, \ and\
  \bibinfo {author} {\bibfnamefont {J.}~\bibnamefont {Hun~Lee}},\ }\href
  {\doibase 10.3938/jkps.61.713} {\bibfield  {journal} {\bibinfo  {journal} {J.
  Korean Phys. Soc.}\ }\textbf {\bibinfo {volume} {61}},\ \bibinfo {pages}
  {713} (\bibinfo {year} {2012})},\ \Eprint {http://arxiv.org/abs/1203.6343}
  {arXiv:1203.6343 [hep-th]} \BibitemShut {NoStop}%
\bibitem [{\citenamefont {Chakrabortty}(2011)}]{Chakrabortty:2011sp}%
  \BibitemOpen
  \bibfield  {author} {\bibinfo {author} {\bibfnamefont {S.}~\bibnamefont
  {Chakrabortty}},\ }\href {\doibase 10.1016/j.physletb.2011.09.112} {\bibfield
   {journal} {\bibinfo  {journal} {Phys. Lett.}\ }\textbf {\bibinfo {volume}
  {B705}},\ \bibinfo {pages} {244} (\bibinfo {year} {2011})},\ \Eprint
  {http://arxiv.org/abs/1108.0165} {arXiv:1108.0165 [hep-th]} \BibitemShut
  {NoStop}%
\bibitem [{\citenamefont {Lee}\ \emph {et~al.}(2009)\citenamefont {Lee},
  \citenamefont {Park},\ and\ \citenamefont {Sin}}]{Lee:2009bya}%
  \BibitemOpen
  \bibfield  {author} {\bibinfo {author} {\bibfnamefont {B.-H.}\ \bibnamefont
  {Lee}}, \bibinfo {author} {\bibfnamefont {C.}~\bibnamefont {Park}}, \ and\
  \bibinfo {author} {\bibfnamefont {S.-J.}\ \bibnamefont {Sin}},\ }\href
  {\doibase 10.1088/1126-6708/2009/07/087} {\bibfield  {journal} {\bibinfo
  {journal} {JHEP}\ }\textbf {\bibinfo {volume} {07}},\ \bibinfo {pages} {087}
  (\bibinfo {year} {2009})},\ \Eprint {http://arxiv.org/abs/0905.2800}
  {arXiv:0905.2800 [hep-th]} \BibitemShut {NoStop}%
\bibitem [{\citenamefont {Chakrabortty}\ and\ \citenamefont
  {Dey}(2016)}]{Chakrabortty:2016xcb}%
  \BibitemOpen
  \bibfield  {author} {\bibinfo {author} {\bibfnamefont {S.}~\bibnamefont
  {Chakrabortty}}\ and\ \bibinfo {author} {\bibfnamefont {T.~K.}\ \bibnamefont
  {Dey}},\ }\href {\doibase 10.1007/JHEP05(2016)094} {\bibfield  {journal}
  {\bibinfo  {journal} {JHEP}\ }\textbf {\bibinfo {volume} {05}},\ \bibinfo
  {pages} {094} (\bibinfo {year} {2016})},\ \Eprint
  {http://arxiv.org/abs/1602.04761} {arXiv:1602.04761 [hep-th]} \BibitemShut
  {NoStop}%
\bibitem [{\citenamefont {Bhattacharya}\ \emph {et~al.}(2013)\citenamefont
  {Bhattacharya}, \citenamefont {Nozaki}, \citenamefont {Takayanagi},\ and\
  \citenamefont {Ugajin}}]{Bhattacharya:2012mi}%
  \BibitemOpen
  \bibfield  {author} {\bibinfo {author} {\bibfnamefont {J.}~\bibnamefont
  {Bhattacharya}}, \bibinfo {author} {\bibfnamefont {M.}~\bibnamefont
  {Nozaki}}, \bibinfo {author} {\bibfnamefont {T.}~\bibnamefont {Takayanagi}},
  \ and\ \bibinfo {author} {\bibfnamefont {T.}~\bibnamefont {Ugajin}},\ }\href
  {\doibase 10.1103/PhysRevLett.110.091602} {\bibfield  {journal} {\bibinfo
  {journal} {Phys. Rev. Lett.}\ }\textbf {\bibinfo {volume} {110}},\ \bibinfo
  {pages} {091602} (\bibinfo {year} {2013})},\ \Eprint
  {http://arxiv.org/abs/1212.1164} {arXiv:1212.1164 [hep-th]} \BibitemShut
  {NoStop}%
\bibitem [{\citenamefont {Kim}\ and\ \citenamefont
  {Hun~Lee}(2016)}]{Kim:2015rvu}%
  \BibitemOpen
  \bibfield  {author} {\bibinfo {author} {\bibfnamefont {N.}~\bibnamefont
  {Kim}}\ and\ \bibinfo {author} {\bibfnamefont {J.}~\bibnamefont {Hun~Lee}},\
  }\href {\doibase 10.3938/jkps.69.623} {\bibfield  {journal} {\bibinfo
  {journal} {J. Korean Phys. Soc.}\ }\textbf {\bibinfo {volume} {69}},\
  \bibinfo {pages} {623} (\bibinfo {year} {2016})},\ \Eprint
  {http://arxiv.org/abs/1512.02816} {arXiv:1512.02816 [hep-th]} \BibitemShut
  {NoStop}%
\bibitem [{\citenamefont {Park}\ and\ \citenamefont {Lee}()}]{Park:2020nvo}%
  \BibitemOpen
  \bibfield  {author} {\bibinfo {author} {\bibfnamefont {C.}~\bibnamefont
  {Park}}\ and\ \bibinfo {author} {\bibfnamefont {J.~H.}\ \bibnamefont {Lee}},\
  }\href@noop {} {\ }\Eprint {http://arxiv.org/abs/2008.04507}
  {arXiv:2008.04507 [hep-th]} \BibitemShut {NoStop}%
\bibitem [{\citenamefont {Park}\ \emph {et~al.}(2022)\citenamefont {Park},
  \citenamefont {Hwang}, \citenamefont {Cho},\ and\ \citenamefont
  {Kim}}]{Park:2022fqy}%
  \BibitemOpen
  \bibfield  {author} {\bibinfo {author} {\bibfnamefont {C.}~\bibnamefont
  {Park}}, \bibinfo {author} {\bibfnamefont {C.-O.}\ \bibnamefont {Hwang}},
  \bibinfo {author} {\bibfnamefont {K.}~\bibnamefont {Cho}}, \ and\ \bibinfo
  {author} {\bibfnamefont {S.-J.}\ \bibnamefont {Kim}},\ }\href@noop {} {\
  (\bibinfo {year} {2022})},\ \Eprint {http://arxiv.org/abs/2205.04445}
  {arXiv:2205.04445 [hep-th]} \BibitemShut {NoStop}%
\bibitem [{\citenamefont {Park}(2015)}]{Park:2015afa}%
  \BibitemOpen
  \bibfield  {author} {\bibinfo {author} {\bibfnamefont {C.}~\bibnamefont
  {Park}},\ }\href {\doibase 10.1103/PhysRevD.91.126003} {\bibfield  {journal}
  {\bibinfo  {journal} {Phys. Rev. D}\ }\textbf {\bibinfo {volume} {91}},\
  \bibinfo {pages} {126003} (\bibinfo {year} {2015})},\ \Eprint
  {http://arxiv.org/abs/1501.02908} {arXiv:1501.02908 [hep-th]} \BibitemShut
  {NoStop}%
\bibitem [{\citenamefont {Park}\ and\ \citenamefont
  {Lee}(2021)}]{Park:2021tpz}%
  \BibitemOpen
  \bibfield  {author} {\bibinfo {author} {\bibfnamefont {C.}~\bibnamefont
  {Park}}\ and\ \bibinfo {author} {\bibfnamefont {J.~H.}\ \bibnamefont {Lee}},\
  }\href@noop {} {\  (\bibinfo {year} {2021})},\ \Eprint
  {http://arxiv.org/abs/2102.06097} {arXiv:2102.06097 [hep-th]} \BibitemShut
  {NoStop}%
\bibitem [{\citenamefont {Park}(2022)}]{Park:2022mxj}%
  \BibitemOpen
  \bibfield  {author} {\bibinfo {author} {\bibfnamefont {C.}~\bibnamefont
  {Park}},\ }\href@noop {} {\  (\bibinfo {year} {2022})},\ \Eprint
  {http://arxiv.org/abs/2209.07721} {arXiv:2209.07721 [hep-th]} \BibitemShut
  {NoStop}%
\bibitem [{\citenamefont {Georgiou}\ and\ \citenamefont
  {Zoakos}(2022)}]{Georgiou:2022ekc}%
  \BibitemOpen
  \bibfield  {author} {\bibinfo {author} {\bibfnamefont {G.}~\bibnamefont
  {Georgiou}}\ and\ \bibinfo {author} {\bibfnamefont {D.}~\bibnamefont
  {Zoakos}},\ }\href@noop {} {\  (\bibinfo {year} {2022})},\ \Eprint
  {http://arxiv.org/abs/2209.14661} {arXiv:2209.14661 [hep-th]} \BibitemShut
  {NoStop}%
\bibitem [{\citenamefont {Park}(2020)}]{Park:2020jio}%
  \BibitemOpen
  \bibfield  {author} {\bibinfo {author} {\bibfnamefont {C.}~\bibnamefont
  {Park}},\ }\href {\doibase 10.1103/PhysRevD.101.126006} {\bibfield  {journal}
  {\bibinfo  {journal} {Phys. Rev. D}\ }\textbf {\bibinfo {volume} {101}},\
  \bibinfo {pages} {126006} (\bibinfo {year} {2020})},\ \Eprint
  {http://arxiv.org/abs/2004.08020} {arXiv:2004.08020 [hep-th]} \BibitemShut
  {NoStop}%
\bibitem [{\citenamefont {Park}(2021)}]{Park:2021wep}%
  \BibitemOpen
  \bibfield  {author} {\bibinfo {author} {\bibfnamefont {C.}~\bibnamefont
  {Park}},\ }\href@noop {} {\  (\bibinfo {year} {2021})},\ \Eprint
  {http://arxiv.org/abs/2106.05500} {arXiv:2106.05500 [hep-th]} \BibitemShut
  {NoStop}%
\bibitem [{\citenamefont {Gubser}(2006)}]{Gubser:2006bz}%
  \BibitemOpen
  \bibfield  {author} {\bibinfo {author} {\bibfnamefont {S.~S.}\ \bibnamefont
  {Gubser}},\ }\href {\doibase 10.1103/PhysRevD.74.126005} {\bibfield
  {journal} {\bibinfo  {journal} {Phys. Rev. D}\ }\textbf {\bibinfo {volume}
  {74}},\ \bibinfo {pages} {126005} (\bibinfo {year} {2006})},\ \Eprint
  {http://arxiv.org/abs/hep-th/0605182} {arXiv:hep-th/0605182} \BibitemShut
  {NoStop}%
\bibitem [{\citenamefont {Park}(2013)}]{Park:2012lzs}%
  \BibitemOpen
  \bibfield  {author} {\bibinfo {author} {\bibfnamefont {C.}~\bibnamefont
  {Park}},\ }\href {\doibase 10.1155/2013/389541} {\bibfield  {journal}
  {\bibinfo  {journal} {Adv. High Energy Phys.}\ }\textbf {\bibinfo {volume}
  {2013}},\ \bibinfo {pages} {389541} (\bibinfo {year} {2013})},\ \Eprint
  {http://arxiv.org/abs/1209.0842} {arXiv:1209.0842 [hep-th]} \BibitemShut
  {NoStop}%
\bibitem [{\citenamefont {Wolf}\ \emph {et~al.}(2008)\citenamefont {Wolf},
  \citenamefont {Verstraete}, \citenamefont {Hastings},\ and\ \citenamefont
  {Cirac}}]{Wolf:2007tdq}%
  \BibitemOpen
  \bibfield  {author} {\bibinfo {author} {\bibfnamefont {M.~M.}\ \bibnamefont
  {Wolf}}, \bibinfo {author} {\bibfnamefont {F.}~\bibnamefont {Verstraete}},
  \bibinfo {author} {\bibfnamefont {M.~B.}\ \bibnamefont {Hastings}}, \ and\
  \bibinfo {author} {\bibfnamefont {J.~I.}\ \bibnamefont {Cirac}},\ }\href
  {\doibase 10.1103/PhysRevLett.100.070502} {\bibfield  {journal} {\bibinfo
  {journal} {Phys. Rev. Lett.}\ }\textbf {\bibinfo {volume} {100}},\ \bibinfo
  {pages} {070502} (\bibinfo {year} {2008})},\ \Eprint
  {http://arxiv.org/abs/0704.3906} {arXiv:0704.3906 [quant-ph]} \BibitemShut
  {NoStop}%
\bibitem [{\citenamefont {Molina-Vilaplana}\ and\ \citenamefont
  {Sodano}(2011)}]{Molina-Vilaplana:2011ydi}%
  \BibitemOpen
  \bibfield  {author} {\bibinfo {author} {\bibfnamefont {J.}~\bibnamefont
  {Molina-Vilaplana}}\ and\ \bibinfo {author} {\bibfnamefont {P.}~\bibnamefont
  {Sodano}},\ }\href {\doibase 10.1007/JHEP10(2011)011} {\bibfield  {journal}
  {\bibinfo  {journal} {JHEP}\ }\textbf {\bibinfo {volume} {10}},\ \bibinfo
  {pages} {011} (\bibinfo {year} {2011})},\ \Eprint
  {http://arxiv.org/abs/1108.1277} {arXiv:1108.1277 [quant-ph]} \BibitemShut
  {NoStop}%
\bibitem [{\citenamefont {Casini}\ and\ \citenamefont
  {Huerta}(2007)}]{Casini:2006es}%
  \BibitemOpen
  \bibfield  {author} {\bibinfo {author} {\bibfnamefont {H.}~\bibnamefont
  {Casini}}\ and\ \bibinfo {author} {\bibfnamefont {M.}~\bibnamefont
  {Huerta}},\ }\href {\doibase 10.1088/1751-8113/40/25/S57} {\bibfield
  {journal} {\bibinfo  {journal} {J. Phys. A}\ }\textbf {\bibinfo {volume}
  {40}},\ \bibinfo {pages} {7031} (\bibinfo {year} {2007})},\ \Eprint
  {http://arxiv.org/abs/cond-mat/0610375} {arXiv:cond-mat/0610375} \BibitemShut
  {NoStop}%
\bibitem [{\citenamefont {Rangamani}\ and\ \citenamefont
  {Takayanagi}(2017)}]{Rangamani:2016dms}%
  \BibitemOpen
  \bibfield  {author} {\bibinfo {author} {\bibfnamefont {M.}~\bibnamefont
  {Rangamani}}\ and\ \bibinfo {author} {\bibfnamefont {T.}~\bibnamefont
  {Takayanagi}},\ }\href {\doibase 10.1007/978-3-319-52573-0} {\emph {\bibinfo
  {title} {{Holographic Entanglement Entropy}}}},\ Vol.\ \bibinfo {volume}
  {931}\ (\bibinfo  {publisher} {Springer},\ \bibinfo {year} {2017})\ \Eprint
  {http://arxiv.org/abs/1609.01287} {arXiv:1609.01287 [hep-th]} \BibitemShut
  {NoStop}%
\end{thebibliography}%

\end{document}